\documentclass[pra,showpacs,twocolumn,superscriptaddress,floatfix,aps,nofootinbib]{revtex4-2}
\usepackage{amsmath,amssymb,amsfonts,bm}
\usepackage{braket}
\usepackage{float}
\usepackage{graphicx}
\usepackage{physics}
\usepackage{mathtools}
\usepackage{hyperref}
\usepackage{comment}
\usepackage[normalem]{ulem}
\usepackage[T1]{fontenc}
\hypersetup{colorlinks,%
    linkcolor=blue,%
    citecolor=blue,%
    urlcolor=blue}
\usepackage{color}
\usepackage{tikz}
\usetikzlibrary{shapes}

\begin{document}
\title{Energy spectra and fluxes of two-dimensional turbulent quantum droplets }
\author{Shawan Kumar Jha}
\affiliation{Department of Physics, Indian Institute of Technology Guwahati, Guwahati 781039, Assam, India}


\author{Mahendra K. Verma}

\affiliation{Department of Physics, Indian Institute of Technology Kanpur 208016, Uttar Pradesh, India}
\author{S. I. Mistakidis}

\affiliation{Department of Physics,Missouri University of Science and Technology, 1315 N. Pine Street, Rolla, MO 65409, USA}

\author{Pankaj Kumar Mishra}
\affiliation{Department of Physics, Indian Institute of Technology Guwahati, Guwahati 781039, Assam, India}

\begin{abstract}
We explore the energy spectra and associated fluxes of turbulent two-dimensional quantum droplets subjected to a rotating paddling potential which is removed after a few oscillation periods. 
A systematic analysis on the impact of the characteristics (height and velocity) of the rotating potential and the droplet atom number reveals the emergence of different dynamical response regimes. 
These are classified by utilizing the second-order sign correlation function and the ratio of incompressible versus compressible kinetic energies. 
They involve, vortex configurations ranging from vortex dipoles to vortex clusters and randomly distributed vortex-antivortex pairs. 
The incompressible kinetic energy spectrum features Kolmogorov ($k^{-5/3}$) and Vinen like ($k^{-1}$) scaling in the infrared regime, while a $k^{-3}$ decay in the ultraviolet captures the presence of vortices. 
The compressible spectrum shows  $k^{-3/2}$ scaling within the infrared and $k$ power law in the case of enhanced sound-wave emission suggesting thermalization. 
Significant distortions are observed in the droplet periphery in the presence of a harmonic trap.  
A direct energy cascade (from large to small length scales) is mainly identified through the flux. 
Our findings offer insights into the turbulent response of exotic phases-of-matter, featuring quantum fluctuations, and may inspire investigations aiming to unravel self-similar nonequilibrium dynamics.

\end{abstract}
\maketitle

\section{Introduction}

Turbulence is a genuine nonequilibrium manifestation of the interplay between order and disorder. 
It arises in a plethora of research fields ranging from magnetohydrodynamics~\cite{beresnyak2019turbulence}, to astrophysics~\cite{kosowsky2002gravitational}, atmosphere~\cite{wyngaard2010turbulence}, non-linear optics~\cite{picozzi2014optical}, and atomic physics~\cite{tsatsos2016quantum}.  
Ultracold atoms are flexible platforms to explore quantum turbulence owing to their exquisite  tunability in terms of system parameters~\cite{bloch2008many,gross2017quantum}. 
Here, the presence of quantum vortices,  being distributed in the quantum fluid known as a Bose-Einstein condensate (BEC), together with their interactions enable a clear differentiation with classical turbulence~\cite{vinen2007quantum,tsubota2008quantum,Bradley2012}.

A key question in the study of turbulence is how energy is transferred across different length scales, defining a direct (inverse) energy cascade from large (small) to small (large) length scales~\cite{frisch1995turbulence,kraichnan1967inertial,kolmogorov1941local}. 
These cascades depend on the system's dimensionality especially in classical turbulence~\cite{kellay2002two,chen2006physical}. 
Indeed, in three-dimensions (3D) there is typically a direct cascade while in two-dimensions (2D) an inverse cascade may  occur~\cite{frisch1995turbulence} originally predicted by Onsager’s model of point-like vortices~\cite{onsager1949statistical}. 
When the system is dissipationless it eventually reaches a quasi-steady state dictated by the well-known (in classical turbulence) Kolmogorov scaling~\cite{kolmogorov1941local}. 
The latter has been observed both experimentally~\cite{Henn:PRL2009, Seman:LPL2011, Thompson:LPL2013, Madeira2020} and numerically~\cite{Kobayashi:PRL2005, Kobayashi:PRA2007} in 3D turbulent BECs 
which also revealed response regimes~\cite{Kobayashi2007, Barenghi:AVSQS2023} absent in their classical counterparts.   

Moreover, unlike classical turbulence the nature of the energy cascade in BECs was shown to depend crucially on the system microscopic properties, e.g., initial conditions, forcing, and interactions~\cite{Sivakumar:POF2024}. 
As an example, it was demonstrated that decaying 2D turbulence in a BEC~\cite{Numasato:PRA2010} supports a direct energy cascade along with Kolmogorov scaling ($k^{-5/3}$) in the incompressible kinetic energy spectra. Also, inverse cascades in 2D were found to arise when large vortex clusters form~\cite{White:PRA2012, Reeves:PRA2014, Stagg:PRA2015, Groszek:PRA2016, Yu:PRA2016, Gauthier:Sci2019}. 
Additionally, Vinen-like scaling ($k^{-1}$) has been reported for fast rotating 2D BECs~\cite{Estrada:PRA2022, Estrada:AVSQS2022} and decaying 2D turbulence originating from the breaking of vortices into multiple ones~\cite{Cidrim2017, Marino2021}. 
Interestingly, in 2D homogeneous superfluids~\cite{Bradley:PRX2012} the presence of a universal ($k^{-3}$) scaling was established in the incompressible kinetic energy spectra at large wavenumbers, whilst a dependence solely on the vortex configuration was showcased at small wavenumbers. 
For completeness, we remark that such scaling behaviors have been reported in binary miscible bosonic settings~\cite{Mithun:PRA2021}, spinor BECs~\cite{kang2017emergence}, and dipolar long-range anisotropically interacting gases~\cite{Sabari:PRA2024,bland2018quantum,bougas2024wave}. 

Another recently realized phase-of-matter appearing in both bosonic mixtures~\cite{Cheiney2018,Semeghini2018,cabrera2018quantum,d2019observation,cavicchioli2024dynamical} and dipolar gases~\cite{bottcher2020new,chomaz2022dipolar} is the so-called incompressible droplet phase with a well-defined surface tension. 
In contrast to BECs, it exists solely in the presence of quantum fluctuations which are usually modeled theoretically with the dimension dependent~\cite{ilg2018dimensional,pelayo2024phases} Lee-Huang-Yang (LHY)~\cite{lee1957eigenvalues} energy correction leading to a suitable extended Gross-Pitaevskii equation (eGPE)~\cite{petrov2015quantum,petrov2016ultradilute}. 
A plethora of droplet properties have been studied, see for instance the reviews~\cite{luo2021new,mistakidis2023few}. 
Of particular interest here, is their capability to coexist with nonlinear excitations such as solitary waves in one-dimension~\cite{katsimiga2023solitary,kopycinski2024propagation,Edmonds_sol} and vortices in 2D~\cite{tengstrand2019rotating,Li:PRA2018,Yang:NJOP2024,Bougas_stability,cheng2024dynamics,Guilong_2024}. 
A few notable examples, in this context, are the ability of higher charge vortices to stabilize when embedded into a droplet background~\cite{Li:PRA2018,Yang:NJOP2024} in sharp contrast to BEC environments, the delay of the eponymous snake instability resulting in vortex generation~\cite{Bougas_stability} and the generic stability of kink configurations in higher-dimensions~\cite{mistakidis2024generic}. 
These are only a few instances where enriched mechanisms as compared to traditional BEC environments have been demonstrated.  
Here, we aim to examine vortex turbulence in a 2D droplet background which, to the best of our knowledge, has not yet been tackled. A particular focus will be placed on 
the underlying kinetic energy spectra and different dynamical response regimes encountered in droplet environments.

To address turbulence in 2D quantum droplet environments, modeled by the appropriate 2D eGPE~\cite{petrov2016ultradilute}, we exploit a stirring potential triggering vortex and sound-wave nucleation. 
As initial states, we calculate a homogeneous droplet environment (suffering from finite-size effects) within a 2D box and a flat-top harmonically trapped configuration. 
Starting from these structures, in the presence of the barrier enforcing a density dip at its location, we let the barrier to stir in the droplet background for three full oscillation cycles and subsequently remove it.

We find that the turbulent response is dictated by the properties of the stirring potential, namely its height and velocity. 
Our analysis based on the second-order sign correlation function, the spectra of the incompressible and compressible kinetic energies and their associated fluxes indicate the appearance of three main response regimes. 
These are dominated by i) vortex dipoles for relatively small velocities and heights of the stirrer, ii) randomly distributed vortex clusters together with an enhanced amount of sound-waves for increasing height and iii) vortex-antivortex clustering for increasing stirring velocity and height of the potential.   
In all cases, the incompressible (compressible) kinetic energy spectra feature a $k^{-3}$ ($k^{-3/2}$) scaling in the ultraviolet (infrared) regime stemming from the presence of vortices. 
Moreover, when vortex dipoles or vortex-antivortex clusters occur the incompressible spectra show Kolmogorov $k^{-5/3}$ scaling in the infrared~\cite{Bradley:PRA2022}, while for random vortex distribution they scale as $k^{-1}$ also known as Vinen scaling~\cite{Baggaley:PRB2012}. 
Conversely, within the ultraviolet regime, the appearance of vortex dipoles is associated with a $k^{-7/2}$ scaling, otherwise a $k$ scaling is observed suggesting a tendency to thermalization. 

For a flat-top droplet background, we identify a transition from a vortex-dipole dynamical regime to one with prevailing vortex-antivortex pairs for increasing stirring velocity. 
Their corresponding energy spectra show a similar behavior to the homogeneous case.  
Interestingly, the droplet boundary becomes substantially deformed due to appreciable density disturbances and vortices generated from the stirring and traveling towards the edges. 
In most of the cases, the accompanied energy fluxes indicate a direct energy cascade, they are enhanced during the second stirring period and suppressed in the third one where the stirring terminates.

The structure of this work proceeds as follows. In Sec.~\ref{sec:MFmodel} we present the 2D eGPE and the stirring protocol used to drive the nonequilibrium  droplet dynamics. 
Section~\ref{sec:spectrum_fluxes} introduces the notion of the energy spectra and fluxes to characterize turbulence. 
The ground states of the box trapped droplet in the presence of the potential barrier for different atom numbers are discussed in Sec.~\ref{ground_state}. 
The emergent turbulent response of the homogeneous box trapped 2D droplet is analyzed in Sec.~\ref{results}, while the dynamics of a harmomically trapped flat-top droplet is presented in Sec.~\ref{turbulence_FT}. 
Finally in Sec.~\ref{sec:con} we conclude and elaborate on future research directions based on our results.

\section{Two dimensional droplet and stirring potential}\label{sec:MFmodel}

We consider a 2D droplet system composed by a homonuclear bosonic mixture experiencing intracomponent repulsion $g_{ \uparrow \uparrow}=g_{ \downarrow \downarrow} \equiv g>0$ and intercomponent attraction of strength $g_{ \uparrow \downarrow}<0$. 
It is trapped in a 2D symmetric box potential of length $L_x=L_y\equiv L$ across the $x$-$y$ plane, while the motion along the tightly confined transverse $z$-direction is frozen~\cite{hadzibabic2011two,huh2024universality}. Experimentally, such a droplet setting may be realized with the hyperfine states $\ket{ \uparrow} \equiv \ket{F=1, m_F=-1}$, $\ket{\downarrow} \equiv \ket{F=1, m_F=0}$ of $^{39}$K~\cite{Cheiney2018,Semeghini2018, Ferioli2019}, whose interactions can be tuned by means of the respective 3D scattering lengths via Feshbach resonances~\cite{Cheiney2018}. 
Specifically, we operate in the regime where the average intracomponent repulsion exceeds the interacomponent attraction, i.e. $\delta g \equiv g_{\uparrow \downarrow}+\sqrt{g_{\uparrow \uparrow} g_{\downarrow \downarrow}} \lesssim 0$, hence entering the droplet region~\cite{luo2021new}. 

Under the above-described assumptions, the description of the corresponding two-component bosonic setting breaks down to an effective single-component eGPE~\cite{petrov2016ultradilute}. The latter is valid for the low-lying droplet excitations~\cite{petrov2015quantum} and its  predictions were confirmed in 3D~\cite{cabrera2018quantum}, see e.g. also Refs.~\cite{flynn2023quantum,Englezos_twocomp_drop,mistakidis2021formation,kartashov2025double} for droplet features beyond the single-component scenario.  
The respective 2D dimensionless eGPE reads  
\begin{equation}
i \frac{\partial \psi}{\partial t} = -\frac{1}{2} \nabla^2 \psi +V_S(t)\psi +  |\psi|^2 \psi \ln\left(|\psi|^2\right), 
\label{eq:2Degpe}
\end{equation}
where $\psi \equiv \psi (x,y)$ and $V_S(t)\equiv V_S(x(t),y(t))$ models an external time-dependent potential (see the discussion below). 
The last logarithmic nonlinear term encompasses both the LHY effects and the mean-field coupling. 
Apparently, at large (low) densities it may become repulsive (attractive). 
Below, all quantities are provided in dimensionless units. 
In particular, the time and length scales are expressed in units of $m/(g n_0 \hbar \sqrt{e})$ and $\sqrt{g n_0 \sqrt{e}}^{-1}$ respectively~\cite{Bougas_stability,mistakidis2024generic} with $n_0$ being the droplet equilibrium density in the thermodynamic limit~\cite{petrov2016ultradilute}. 
The droplet wave function is normalized to the total number of atoms as $\int |\psi|^2~dx dy =gN$.  

The time-dependent external potential term in Eq.~(\ref{eq:2Degpe}) is essential for the type of the induced nonequilibrium dynamics and, in particular,  the turbulent response attained at longer timescales. 
It models a rotating repulsive barrier, dubbed obstacle, which has been intensively used in past  Bose gas experiments~\cite{Kwon:PRA2015,Kwon:PRL2016,Seo:Nature2017} to generate vortex  turbulence. 
Along these lines, it is employed herein to stir  the droplet and drive it into the turbulent regime. It is given by 
\begin{equation}\label{eq:Vobs}
V_{S}(t) = V_0 \exp\left[ -\frac{[x-x_0(t)]^2 + [y-y_0(t)]^2}{\sigma_0^2} \right], 
\end{equation}
where $x_0(t) = r_0 \cos(2\pi t/T_{{\rm osc} })$, $y_0(t) = r_0\,\sin(2\pi\,t/T_{{\rm osc} })$. The rotation radius of the obstacle is $r_0$, $T_{{\rm osc} }$ is the time period of the rotation, while $\sigma_0$ and $V_0$ model the width and height of the obstacle respectively.  Hence, the angular velocity of the obstacle is $v=2\pi r_0/T_{{\rm osc} }$.

In the following, we analyze the emergent vortex turbulence emanating from the rotation of the gaussian obstacle inside the 2D droplet environment. 
Specifically, we first obtain the ground state of the 2D droplet in the presence of the static gaussian obstacle placed at $(r_0, 0)$ by solving the time-independent eGPE of Eq.~(\ref{eq:2Degpe}). 
This is achieved via the imaginary-time propagation technique incorporating the split-step Fourier method for the time derivatives~\cite{Jha:2023}. 
Once the ground state of the 2D is obtained the obstacle starts to rotate  within the droplet and we monitor the dynamics of the system using the real-time propagation scheme.

Depending on the system parameters, and especially of the obstacle, the rotation seeds vortex dipoles whose number and distribution are dictated by the obstacle characteristics~\cite{Reeves:PRA2012}. 
Here, we conduct different simulations characterized by a varying height ($V_0$) of the stirring potential and a fixed width $\sigma_0=L/20 = 3$. The latter is sufficiently small to ensure the generation of defects. 
We choose to rotate the obstacle for three complete periods which suffice to generate vortices inside the droplet and afterwards it is switched off. 
Once the stirring protocol is terminated we analyze the nonequilibrium  dynamics of the vortices and the subsequent turbulent response. 
For all of our simulations to be presented below, we use a 2D box of length $L_x=L_y=60$ with spatial discretization $dx=dy\approx 0.08$ and periodic boundary conditions. For the time-integrator our timestep is fixed to $dt = 0.0001$. 
These numerical ingredients can ensure the appropriate numerical convergence of our 2D simulations.

\section{Energy spectrum and fluxes}\label{sec:spectrum_fluxes}

The energy transport across different length scales during the nonequilbrium dynamics of the many-body system is crucial for detecting and characterizing the emergent turbulent response. 
Accordingly, it is imperative to inspect the (kinetic) energy spectra and associated flux in order to analyze  the turbulent behaviour of  both classical and quantum fluids~\cite{madeira2020quantum,tsatsos2016quantum}. 
Using the Madelung transformation~\cite{madelung1927quantum}, $\psi(\bm{r}) = \sqrt{\rho(\bm{r})} e^{i\phi(\bm{r})}$, it is possible to define the  density-weighted velocity field 
\begin{equation}
\bm{w}(\bm{r}) = \sqrt{\rho(\bm{r})}\bm{v}(\bm{r}). 
\end{equation}
Here, $\rho(\bm{r})$ is the droplet density, $\bm{v} = \nabla \phi$ refers to the corresponding velocity field and $\phi$ is the phase of the complex  wave function. 
As such, the kinetic energy spectra can be expressed in terms of the Fourier transform of the density-weighted velocity field~\cite{Bradley:PRA2022} as follows
\begin{equation}
\label{eq:spectra_def}
E^{\alpha}_{{\rm kin} }(t) = \frac{1}{2} \int \left|\mathcal{F}_{\bm{k}}\left[\bm{w}^{\alpha}\right]\right|^2 d\bm{k}= \int \varepsilon_{{\rm kin} }^{\alpha}(k)dk.
\end{equation}
In this expression,   $\mathcal{F}_k[.]$ represents the Fourier component of the field, while the index $\alpha\in\{i,c\}$ denotes either the incompressible ($\alpha\equiv i$) or the compressible ($\alpha \equiv c$) kinetic energy contribution. 
The first signifies the appearance of topological defects and the latter signals the rise of acoustic (sound) waves in the system~\cite{Nore1997}. 
These contributions may be identified through the Helmholtz decomposition of the velocity field~\cite{Nore1997,tsatsos2016quantum} and separated into a divergence-free (incompressible) and a curl-free (compressible) part as $\bm{w} = \bm{w}^i + \bm{w}^c$.

In what follows, in order to numerically calculate the high-resolution spectra of the droplet we adopt the methodology outlined in Ref.~\cite{Bradley:PRA2022} utilizing the angle-averaged Wiener-Khinchin theorem. The latter essentially relates the auto-correlation of a field, $\Phi$, with it's power-spectra in fourier space, namely 
\small
\begin{equation}
|\mathcal{F}\left[\Phi(x)\right]|^2 = \mathcal{F}\left[C(x)\right] = \mathcal{F}\left[\int_{-\infty}^{\infty} \Phi^{*}(x)\Phi(x+x')dx'\right].
\end{equation}
\normalsize
Specifically, in Ref.~\cite{Bradley:PRA2022}, an angle-averaged form of the Wiener-Khinchin   theorem was used to obtain the kinetic energy spectra 
\begin{equation}
\label{eq:spectra_khinchin}
\varepsilon_{{\rm kin} }^{\alpha}(k) = \frac{1}{4\pi}\int d^2\bm{r}J_0(k|\bm{r}|)C[\bm{w}^{\alpha},\bm{w}^{\alpha}](\bm{r}),
\end{equation}
where $J_0$ is the 0\textsuperscript{th} order Bessel function of the first kind and $C$ is the auto-correlation function in real space. 
Moreover, the flux corresponding to the incompressible and compressible kinetic energy contributions is obtained using the spectral decomposition~\cite{Sivakumar:POF2024} 
\begin{equation}
\label{eq:flux_def}
\Pi_{{\rm kin} }^{\alpha}(k) = -\frac{d}{dt}\int_{k_0}^{k} \varepsilon^{\alpha}_{{\rm kin} }(k') dk',
\end{equation}
with $k_0 = 2\pi/L$ representing the smallest wavenumber and $L$ being the system size. 
A positive flux implies the existence of a direct energy cascade transporting energy from larger to smaller length scales in the droplet, while a negative flux represents an inverse energy cascade where energy transports from smaller to larger length scales.

\section{Initial droplet states}\label{ground_state}

\begin{figure}[!htbp]
\centering
\includegraphics[width=\linewidth]{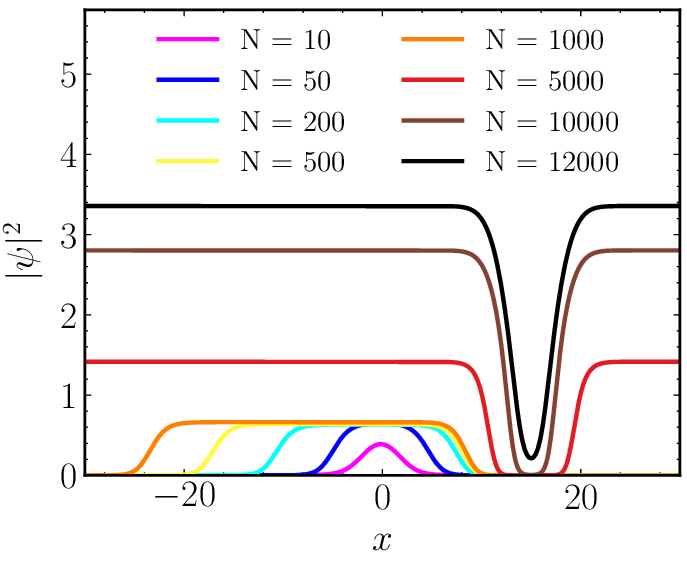}
\caption{Ground state droplet densities along the $x$-direction at $y=0$ for different atom numbers $N$ (see legend) in the presence of a gaussian obstacle placed at $(x_0(0),y_0(0))=(15,0)$. 
The obstacle is characterized by height $V_0 = 5.5$, and width $\sigma_0 = 3$. 
It can be seen that for increasing $N$, the droplet density reshapes from a gaussian profile to a flat-top one (when $N\gtrsim 50$) and afterwards becomes homogeneous due to the  finite size ($L \times L =60 \times 60$) of the box potential. The density dip centered at $x=15$ is due to the presence of the obstacle.}\label{fig:ground_states}
\end{figure}

The droplet in the 2D box with periodic boundary conditions is initiated in its ground state configuration with the gaussian obstacle residing at $(x_0(t=0)=r_0\equiv 15, y_0(t=0)=0)$. 
The corresponding density profiles along the $x$-direction at $y=0$ are presented in Fig.~\ref{fig:ground_states} for various atom numbers, $N$. 
It becomes apparent that for small $N$, e.g. $N=10$, the droplet exhibits a gaussian-type distribution. 
In contrast, an increasing particle number results in the gradual  formation of flat-top droplets whose peak density remains fixed and their width becomes larger, see in particular the cases of $N=50$ up to $N=1000$. This is a manifestation of the incompressible nature of droplets and it is in line with previous predictions~\cite{mistakidis2023few}. 
Importantly, here the droplet configurations containing $N>200$ are not symmetric with respect to $x=0$ due to the gaussian obstacle at $x(t=0)=15$, $y(t=0)=0$ which pushes them along the $x<0$ direction. 
On the other hand, a further increase of the atom number, $N\gtrsim 5000$, leads to a homogeneous droplet background even in the absence of the obstacle as a result of the finite box size\footnote{We remark, here, that the total energy of the system remains negative for $N \leq 9700$.}. 
Due to the involvement of finite size effects, here, the droplet background increases for larger atom number, while the obstacle imprints a density notch at its position on the droplet. Notice that besides the fact that the obstacle characteristics remain the same, the width and amplitude of the aforementioned density notch depend on $N$ which can be traced back to the modification of the droplet background with $N$.

To induce the dynamics, the gaussian obstacle starts to rotate within the droplet around the center of the box at  radius $r_0 = L/4 = 15$. 
In our studies, we consider different heights, $V_0$, and velocities, $v$, of the stirring potential which has fixed width $\sigma_0=L/20 = 3$. 
The latter is, in general, larger than the droplet healing length facilitating the production of vortical defects~\cite{Kwak:PRA2023,Kokubo:JLTP2024}. 
Moreover, we discuss the impact of two different droplet states in the dynamics.  Namely, the case of a large atom number ($N\gtrsim 5000$) where the droplet has an almost homogeneous profile and suffers from finite size effects [Sec.~\ref{results}], and the flat-top (non-uniform) 2D droplet state [Sec.~\ref{turbulence_FT}]. 
In the last case, we assume an obstacle placed within the droplet background.

\section{Vortex Turbulence in a uniform droplet}\label{results}

\subsection{Different turbulent regimes  of the  droplet}\label{QDs states}

\begin{figure*}[!htbp]
\centering
\includegraphics[width=1.0\textwidth]{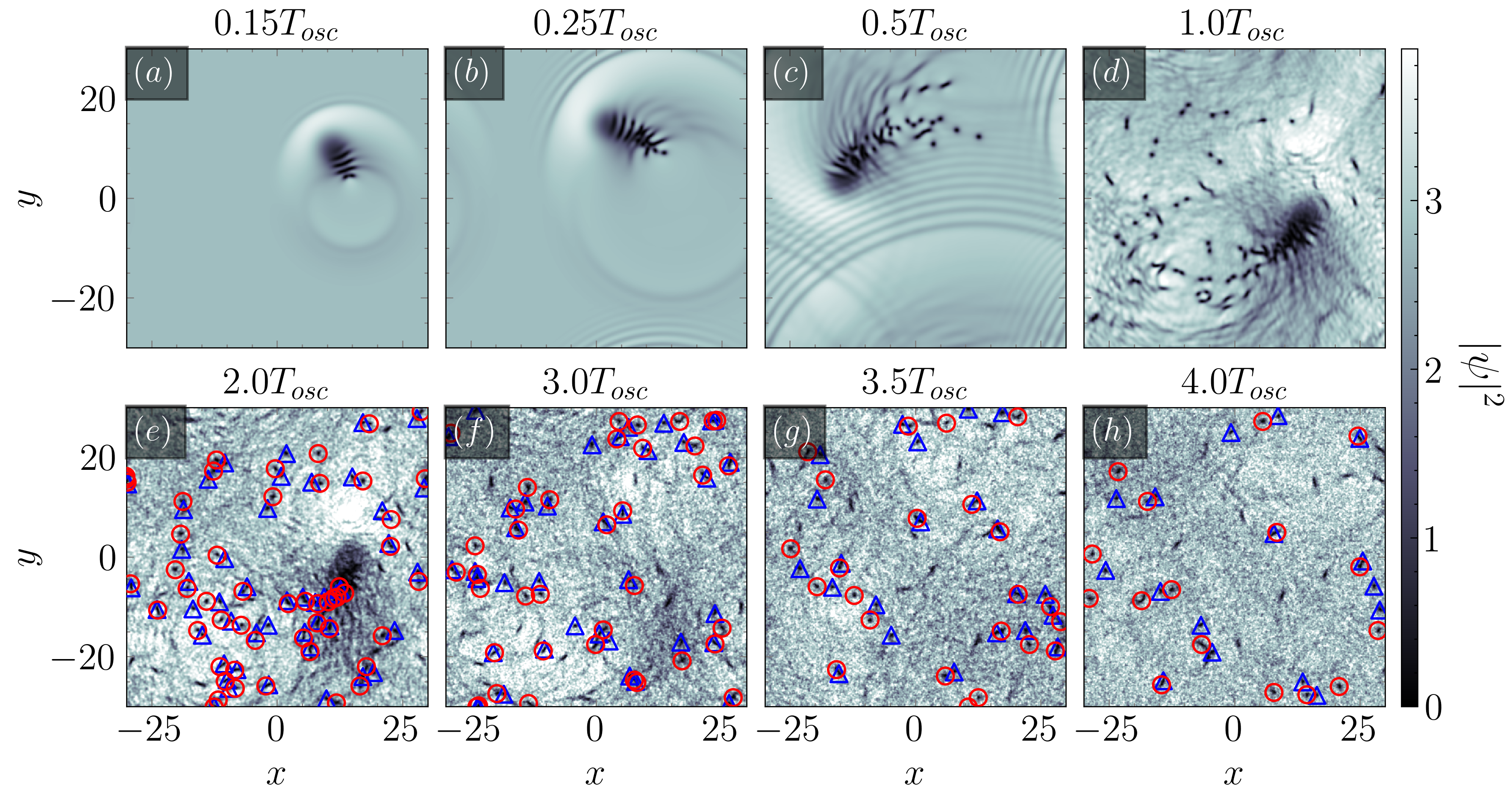}
\caption{Density snapshots (see legends) across the $x$-$y$ plane of a 2D homogeneous droplet environment subjected to a stirring potential. 
The stirrer is rotated for three complete periods within the droplet and it is ramped down at the end of the last period. In the course of the stirring, panels (a)-(f), a large number of vortices is  seeded in the wake of the gaussian obstacle. 
Red circles (blue triangles) in panels (e)-(h) mark the location of vortices (antivortices) identified from the respective wave function phase. 
During the evolution a large amount of vortices are annihilated [panels (c)-(e)] and the remaining ones support the turbulent response,  see panels (f)-(h).  
The droplet consists of $N=10^4$ bosons and the gaussian obstacle given by Eq.~(\ref{eq:Vobs}) is characterized by height $V_0=3.5$, angular velocity $v=2.5$ ($T_{{\rm osc} }\sim37.7$) and width $\sigma_0=3.0$.}\label{fig:density_shedding}
\end{figure*}

To visualize the different stages of the emergent dynamical response,  we first monitor the 2D droplet density in the course of the rotation and after the removal of the gaussian obstacle. 
The latter is characterized here by height $V_0 = 3.5$ and angular velocity $v = 2.5$, while it is rotating for three complete periods. 
The resulting density dynamics of an initial homogeneous droplet is presented in Fig.~\ref{fig:density_shedding}. 
It becomes apparent that the rotating obstacle gives rise to acoustic wave formation at the initial stages of the evolution, see the created density ripples in Fig.~\ref{fig:density_shedding}(b), (c). 
Subsequently, vortex pair generation is observed in the wake of the obstacle\footnote{It is interesting to note that due to the finite box size the nucleated acoustic waves travel towards the edges and then reflected back. This proliferates the wave turbulent component in the dynamics which in an ideal infinite system would be less prominent.}, see e.g. the density dips in Fig.~\ref{fig:density_shedding}(c), (d), similarly to the case of repulsive Bose gases~\cite{gauthier2019giant}. 

The ensuing vortex-antivortex pairs are quantified by the phase of the 2D droplet wave function characterized by a clockwise and counterclockwise $2\pi$ rotation respectively. 
To facilitate the identification of the vortices (antivortices) we mark their locations by red circles (blue triangles) after the first oscillation period of the obstacle illustrated in Fig.~\ref{fig:density_shedding}(e)-(h). The creation of the vortex distribution throughout the droplet, after the first stirring period, is accompanied by an  inevitable coarsening stage. 
Here, vortex annihilation processes when vortex-antivortex pairs come close to each other (meaning that they approach each other at distances smaller than the droplet  healing length) take place. 
This is reflected by the smaller amount of vortices at longer evolution times, compare for instance Fig.~\ref{fig:density_shedding}(f) and (h). 
The characteristics of the vortex distribution leading to this turbulent state along with the existence of the latter are analyzed below with respect to the characteristics of the stirring potential.  

To shed light on the different dynamical response regimes of the droplet, we next explore the type of the vortex distribution preceding the turbulent stage at long evolution times. 
Specifically, we study the impact of the striring characteristics (i.e., velocity and height of the gaussian obstacle) as well the total atom number in the droplet on the vortex distribution. 
To characterize the latter we inspect the type of vortex clustering taking place in the system. 
Vortex clustering has been quantified in repulsive Bose gases using various methods, e.g. based on statistical pattern
recognition methods utilizing the  Ripley's K function~\cite{White:PRA2012}, the so-called recursive clustering algorithm~\cite{Reeves:PRA2014}, the dipole moment of the vortex distribution~\cite{Groszek:PRA2016} or the second order vortex sign correlation function~\cite{ Seo:Nature2017}. 
Here, we deploy the vortex sign correlation function defined as 
\begin{equation}
C_2 = \frac{1}{2N_v}\sum_{i=1}^{N_v}\sum_{j=1}^2 c_{ij}, 
\end{equation}
with $N_v$ denoting the total  number of existing vortices in the system at a specific time-instant. 
Importantly, $c_{ij} = 1$ if the $i$\textsuperscript{th} vortex and it's $j$\textsuperscript{th} nearest vortex have the same charge, otherwise $c_{ij} = 0$~\cite{Seo:Nature2017}. 
It turns out that if there are more vortex dipoles (namely vortex pairs of opposite sign) than clusters of vortices with the same sign present in the system, then $C_2<0.5$. 
On the other hand, $C_2=0.5$ represents a random vortex configuration with equal presence of vortex dipoles and clusters, whilst $C_2>0.5$ implies dominance of vortex clusters. 
\begin{figure*}[!htbp]
\centering
\includegraphics[width=\linewidth]{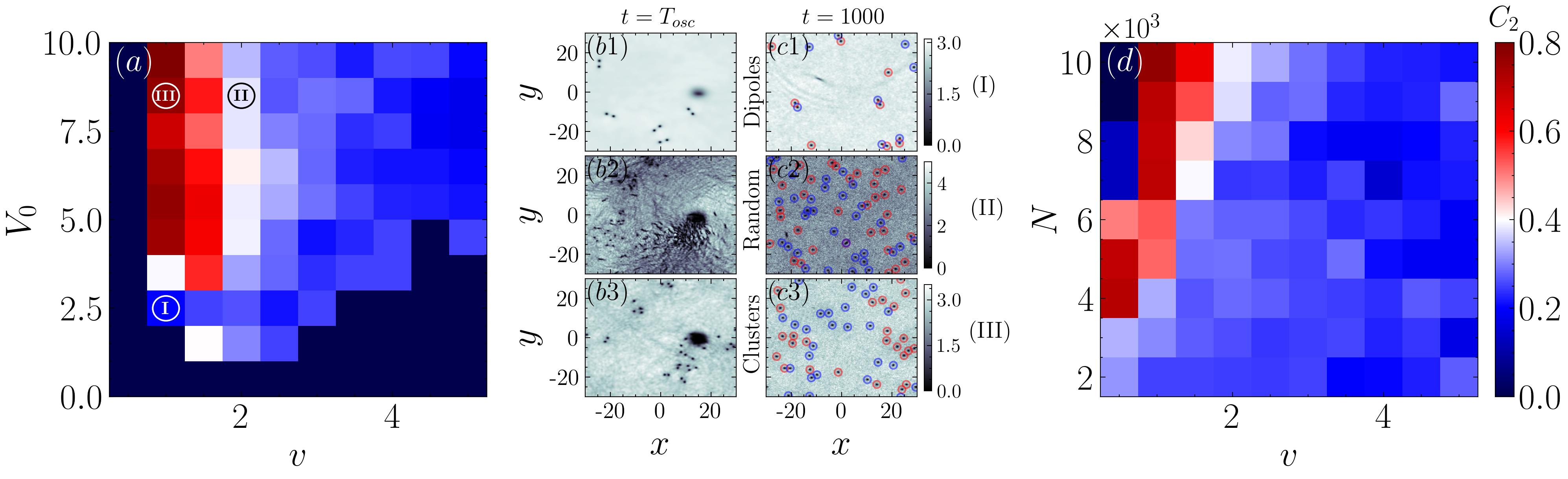}
\caption{
Phase diagram of the vortex sign correlation function ($C_2$) measured after the first oscillation period of the obstacle for distinct values of (a) $V_0$ and $v$  as well as (d) $N$ and $v$. Colormaps are the same for panels (a) and (d). It can be seen that in general vortex dipole distributions are proliferated, see the parametric regions  where $C_2<0.5$. Black regions indicating $C_2 \approx 0$ correspond to cases where vortex shedding is suppressed. Characteristic droplet density profiles at the end of the first stirring period, $t = T_{{\rm osc} }$, and at $t=1000$ are depicted in panels (b1)-(b3) and (c1-c3) respectively for the cases marked as I, II and III in $C_2$ of panel (a). 
Note that in panels (b1), (c1), (b3), (c3) [(b2), (c2)] $T_{{\rm osc}} \sim 94.25$ [$T_{{\rm osc}} \sim 47.12$] and hence  $t=1000\sim10.6T_{osc}$   [$t=1000\sim21.2T_{{\rm osc}}$]. Red (blue) circles in panels (c1)-(c3) indicate vortices (antivortices) exhibiting a clockwise (counter-clockwise) $2 \pi$ phase circulation. }\label{fig:C2_phase_diagram}
\end{figure*}

The vortex sign correlation function after the first oscillation period, is presented in Fig.~\ref{fig:C2_phase_diagram}(a), (d) in the $V_0$-$v$ and $N$-$v$ parametric planes elucidating the complex interplay between height and angular velocity of the gaussian obstacle and the total particle number. 
Here, we calculate $C_2$ at the end of the first stirring period since afterwards the interactions among the gaussian obstacle and the already nucleated vortices become noticeable, along with shedding additional  vortices, thereby disrupting their arrangement. 
In particular, Fig.~\ref{fig:C2_phase_diagram}(a) illustrates the phase diagram of $C_2$ obtained for a set of 100 different simulations with respect to $V_0$ and $v$. 
For relatively small velocities $v<2$, we observe that an increasing height of the stirrer ($V_0>5$), leads to a transition from a vortex dipole where $C_2<0.5$ to a vortex cluster with $C_2>0.5$ dominated distribution. 
Signatures of these distributions  manifest already after the first stirring period. 
To visualize these defect configurations we present in Fig.~\ref{fig:C2_phase_diagram}(b1), (b3) the density of the driven homogeneous droplet background at the end of the first stirring period for $(v,V_0) = (1.0,2.5)$ and $(v,V_0) =(1.0, 8.5)$ respectively. These will be dubbed cases I and III in what follows. 
It can be readily seen that in case I there is a periodic shedding of vortex dipoles from the obstacle which persist for long evolution times, see Fig.~\ref{fig:C2_phase_diagram}(c1). 
However, in case III the stirrer mostly triggers small vortex clusters which occupy the entire background at later times as showcased in Fig.~\ref{fig:C2_phase_diagram}(c3). 

A similar to the above-described  increasing trend of $C_2$ appears at $v=2$ with respect to $V_0$. 
However, here approximately $C_2 \to 0.5$ for $V_0>3$ which implies the formation of a random vortex distribution.  Characteristic densities for $(v, V_0) = (2.0,8.5)$ (named case II) after the first oscillation period and at long times ($t=1000$) are shown in Fig.~\ref{fig:C2_phase_diagram}(b2) and (c2) respectively. 
It becomes evident that the obstacle induces a large amount of acoustic waves and simultaneously sheds a random vortex  distribution in its wake. 
In contrast, when the angular velocity of the obstacle is increased $v>2$, the vortex sign correlation function appears to decrease remaining below 0.5 and hence signifying the gradual prevalence of vortex dipoles in the system.  
Notice here that the black regions in the phase diagram where $C_2\approx 0$ indicate that for these characteristics of the gaussian obstacle it was not possible to identify vortex production in the simulations; an observation that holds true even for longer evolution times in most of these cases.

The competition of the total number of particles $N$ and the angular velocity of the stirrer on the respective vortex distribution captured by $C_2$ is shown in Fig.~\ref{fig:C2_phase_diagram}(d). It can be seen that a larger atom number at relatively small velocities ($1<v<2$) results in a growing behavior of $C_2$ hinting towards enhanced nucleation of  vortex clusters in denser systems. Otherwise, it is clear that an increasing velocity generally facilitates the creation of vortex dipoles. 
Concluding, we can infer that a distribution of vortex dipoles is energetically favorable except from regions of relatively small velocity combined either with high obstacles or large atom numbers which proliferate the generation of vortex clusters.

\begin{figure}[!htbp]
\centering
\includegraphics[width=\linewidth]{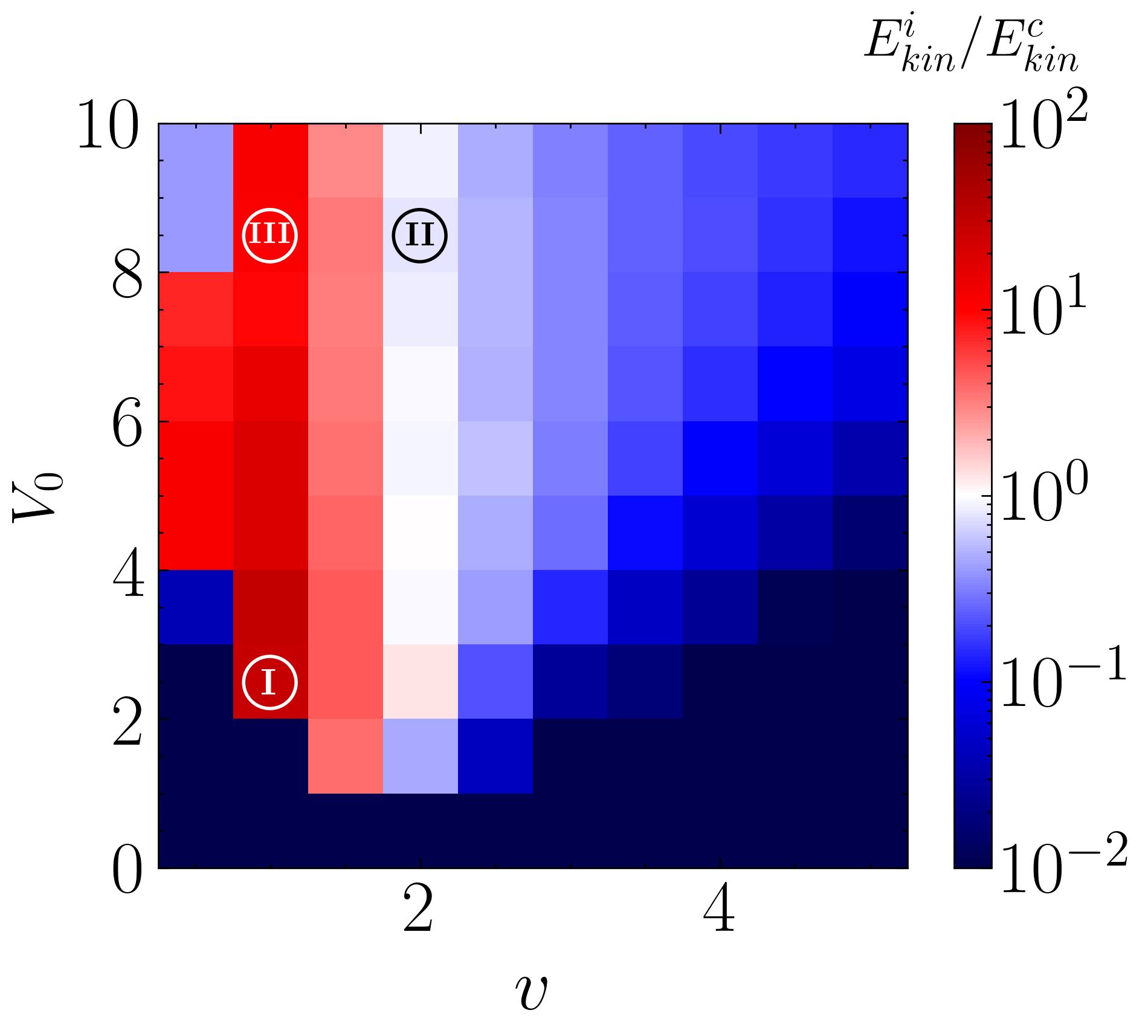}
\caption{Phase diagram of the ratio, $\zeta \equiv E_{kin}^i/E_{kin}^c$,  between the  incompressible and compressible kinetic energy of the droplet environment with respect to the obstacle height $V_0$ and angular velocity $v$ averaged over three oscillation periods, i.e., from $T=0$ to  $T=3T_{{\rm osc} }$.  
In general, for $v<2$ the ratio is large implying that vortical defects are generated and hence vortex turbulence is anticipated. Otherwise, nucleation of acoustic waves dominates and wave turbulence is favorable. 
Cases I, II and III, are characteristic ones (labeled in Fig.~\ref{fig:C2_phase_diagram}) where compressible or incompressible kinetic energy parts dominate having a ratio $\zeta=28.28$, $\zeta=0.80$ and $\zeta=10.97$ respectively.}\label{fig:phase_plot_ekic_ratio}
\end{figure}

Having exemplified the parametric regions where each vortex distribution dominates it is essential to measure the underlying kinetic energy contributions after the stirring process has been terminated and the turbulence state is approached. 
As explained in Sec.~\ref{sec:spectrum_fluxes}, the  incompressible (compressible) kinetic energy part signifies the generation of vortical defects (acoustic waves) and therefore implies prevalence of vortex (wave) turbulence. 
The phase diagram representing the ratio $\zeta = E_{{\rm kin} }^i/E_{{\rm kin} }^c$ of the incompressible ($E_{{\rm kin} }^i$) to the  compressible ($E_{{\rm kin} }^c$) kinetic energy  generated in the homogeneous droplet due to stirring is depicted in Fig.~\ref{fig:phase_plot_ekic_ratio} as a function of the height and the angular velocity of the gaussian obstacle. 
Notice that the parameter $\zeta$ is computed as a temporal average within the first three complete oscillation periods, namely in the interval ($0,3T_{{\rm osc} }$). 
Afterwards, stirring is terminated and defect  annihilation becomes prominent. 

Overall, we can discern that the angular velocity of the stirrer essentially dictates the type of emergent turbulence whose strength can be further regulated by the height of the obstacle. 
Indeed, for $v<2$ vortex turbulence is in general favorable since $\zeta>1$; otherwise wave turbulence is at play. 
Also, an increasing (decreasing) obstacle height promotes vortex (wave) turbulence at the appropriate angular velocity region. 
This is somewhat expected since lower obstacles favor the creation of sound waves in the system rather than topological defects as in the case of repulsive Bose gases~\cite{andrews1997propagation,dalfovo1999theory}.
A maximum value of $\zeta$ is observed for case I, with $\zeta=28.28$,  suggesting that the system's evolution is largely dominated by vortices (and in particular vortex dipoles since $C_2<0.5$ in Fig.~\ref{fig:C2_phase_diagram}(d)) rather than sound waves. 
Remaining in the same velocity region where incompressible kinetic energy prevails one can also realize case III at which  $\zeta=10.97$ associated with an arguably large $C_2 > 0.5$ indicating vortex clustering, see also Fig.~\ref{fig:C2_phase_diagram}(d). 
This suggests pronounced vortex clustering in the system, despite the presence of sound waves (evident in the temporal density which is not shown here for brevity) that hinder such a process as known from repulsive Bose gases~\cite{Kanai:PRL2021}. 
Finally, turning to the case where $\zeta<1$ meaning that compressible kinetic energy prevails, see e.g. case II with $\zeta = 0.80$, production of acoustic waves is substantial in the system. 
The above-mentioned three distinct cases are representative of the kind of turbulence and vortex/sound-wave distribution occurring in the droplet\footnote{We remark that in all cases by monitoring the number of vortices in the course of the evolution we can deduce that during the stirring period it increases almost linearly and afterwards (due to annihilations) a power law decay is observed. This decay is more  prominent in case II.} and will be used in the following for analyzing the underlying energy spectra.

\subsection{Incompressible Spectra and Fluxes}\label{sec:fluxes_homogeneous}

The identification of parametric regions where incompressible and compressible kinetic energy contributions prevail prompt us to subsequently investigate their spectra and associated fluxes.  
These observables are customary to characterize turbulent  flows~\cite{tsatsos2016quantum} and they will enable us to deduce underlying scaling laws at different length scales as well as the existence of direct or inverse energy cascades in the 2D quantum droplet. Below, we consider the characteristic three different cases (I, II and III) corresponding to specific heights and velocities of the obstacle and triggering vortex (cases I, III) and wave turbulence (case II). 
Recall that they have been identified in the vortex sign correlation function [Fig.~\ref{fig:C2_phase_diagram}] and the ratio of incompressible to compressible kinetic energies [Fig.~\ref{fig:phase_plot_ekic_ratio}] as representative ones of distinct turbulent regimes in our driven droplet setting.

\begin{figure*}[!t]
\centering
\includegraphics[width=\linewidth]{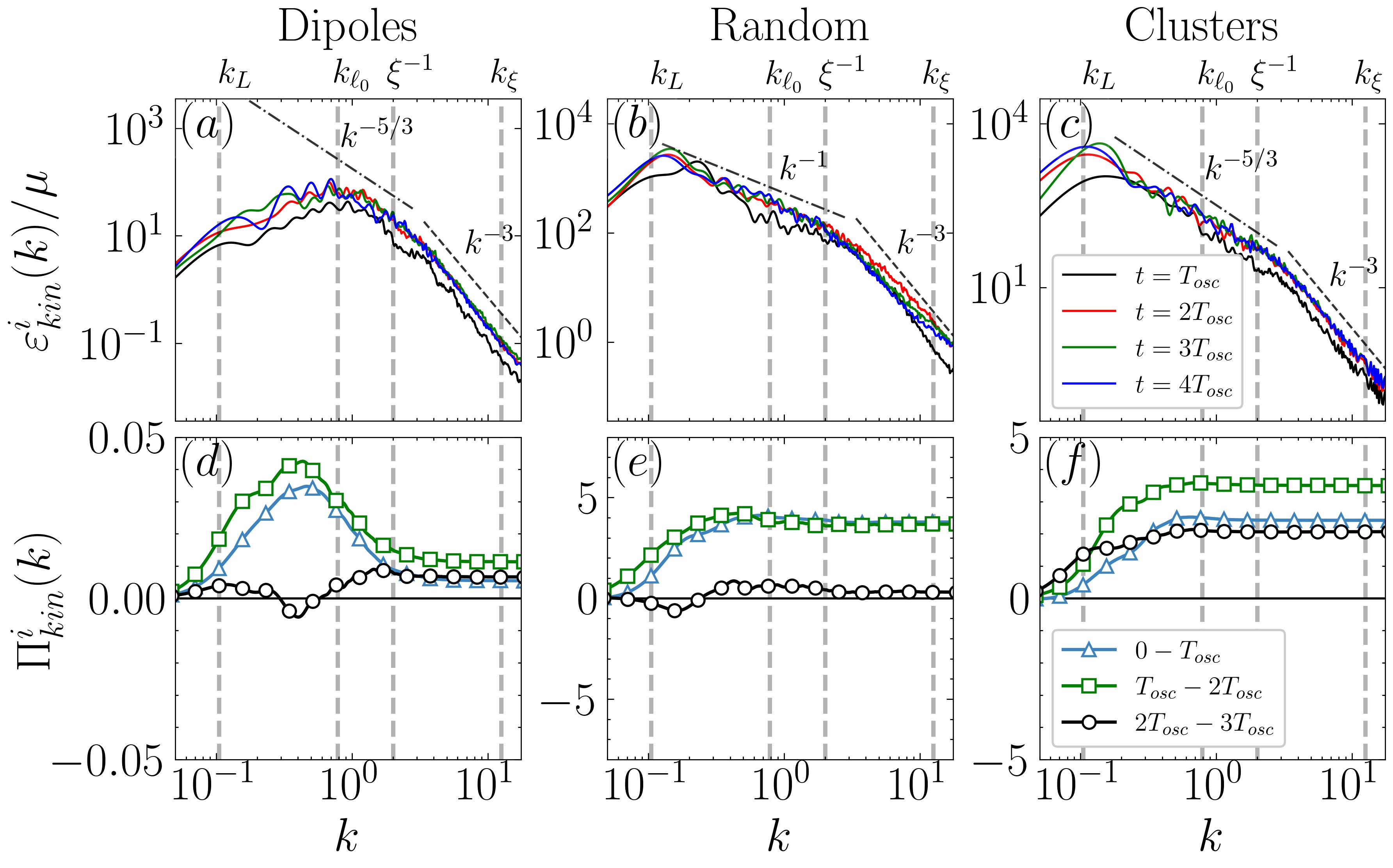}
\caption{(a)-(c) Incompressible kinetic energy spectra and (d)-(f) corresponding energy fluxes of the 2D stirred homogeneous droplet environment of $N=10^4$ at distinct times in terms of the stirring period (see legends). 
The rotating potential is characterized by (a), (d) $(v,V_0) = (1.0,2.5)$ dubbed case I, (b), (e) $(v, V_0) = (2.0,8.5)$ named case II and (c), (f) $(v,V_0) =(1.0, 8.5)$ called case III, see also Fig.~\ref{fig:C2_phase_diagram}(d).  
Here, $k_L$, $k_{l_{0}}$ and $k_{\xi}$ denote the momentum scales associated with the box length, mean intervortex distance and healing length $\xi$ respectively, while $\mu$ is the chemical potential of the droplet background. 
The black solid lines in panels (d)-(f) mark zero flux. 
As shown, a $k^{-3}$ scaling occurs in the ultraviolet ($k>\xi^{-1}$) regime irrespectively of the characteristics of the stirrer.  
Kolmogorov scaling, $k^{-5/3}$, in the infrared regime ($k<\xi^{-1}$) is observed for cases I and III associated with vortex turbulence. 
Notice the arguably larger range of relevant wavenumbers in case III, i.e. $k_{l_0}<k<\xi^{-1}$, where vortex clusters exist as compared to case I, namely $k_{L}<k<\xi^{-1}$, characterized by vortex dipoles. 
In contrast, a $k^{-1}$ scaling takes place in the infrared range for case II where sound-waves dominate. The energy flux remains positive in all cases and it is larger in the second stirring period compared to the first, whilst it is suppressed during the third period at the end of which the gaussian barrier is removed.}\label{fig:EKI_spec_uniform}
\end{figure*}

Let us start with the incompressible kinetic energy spectra ($\varepsilon^{i}_{kin}(k)$) and fluxes ($\Pi^i_{{\rm kin} }(k)$) presented in Fig.~\ref{fig:EKI_spec_uniform} at different oscillation periods ($T_{{\rm osc} }$) of the stirrer and for all three different cases. 
Overall, we observe that the spectra from  larger to smaller wavenumbers exhibit an increasing trend reaching a maximum at a particular momentum (around $k_{l_0}=2 \pi/ l_0$ and $k_L= 2 \pi /L$) depending on the driving characteristics as described by cases I, II and III. 
Here, $L$ and $l_0$ represent the box length and intervortex distance respectively (see also the discussion below). 
The increasing behavior of $\varepsilon^{i}_{{\rm kin} }(k)$ is inherently related to the presence of vortical defects. 
As such, $\varepsilon^{i}_{{\rm kin} }(k)$ acquires a maximum close to $k_{l_0}$ in case I [Fig.~\ref{fig:EKI_spec_uniform}(a)] due to the dominance of vortex dipoles, while it becomes maximum about $k_L$ in case III [Fig.~\ref{fig:EKI_spec_uniform}(c)] since vortex clusters occur. 
Recall that in case II [Fig.~\ref{fig:EKI_spec_uniform}(b)] a random vortex distribution exists captured by $C_2\approx 0.5$, while sound-waves dominate. Hence, $\varepsilon^{i}_{{\rm kin} }(k)$ maximizes around $k_L$.
For smaller momenta, $\varepsilon^{i}_{{\rm kin} }(k)$ experience a descending tendency evidencing the absence of vortical defects. 
The above-described behavior of the incompressible kinetic energy spectra is characteristic of turbulent response as has been also numerically showcased for Bose gases~\cite{Bradley:PRX2012,Nore1997}.

Moreover, it should be noticed that, irrespectively of the driving characteristics (referring to cases I, II and II), the incompressible kinetic energy spectra show a self-similar behavior~\cite{madeira2020quantum,tsatsos2016quantum,bougas2024wave} at large momenta (ultraviolet regime) and long evolution times by means that they collapse one atop the other. 
In this region of large momenta and for all three cases, the spectra feature a $k^{-3}$ scaling at the ultraviolet range, i.e. $k>\xi^{-1}$, with $\xi$ being the healing length of the 2D droplet~\cite{luo2021new,petrov2016ultradilute}. 
This scaling arises due to the structure of the vortex cores similar to the case of a scalar BEC~\cite{Bradley:PRX2012}, see also Appendix~\ref{scaling_vortex} for details. 
Despite these similarities, however, there are also discernible spectral features for the different cases. Indeed, for case I, there is a $k^{-5/3}$ Kolmogorov scaling in between $k_{\ell_0}$ and $\xi^{-1}$ with $\ell_0$ being the mean intervortex distance. 
The latter has been measured numerically using $l_0 = 1/\sqrt{N_v}$ which holds in repulsive Bose gases~\cite{Sivakumar:POF2024}, where $N_v$ is the number of vortices present in the system\footnote{Note that the validity of the used $l_0$ is an open issue for droplets and might be slightly different. However, for our quantitative purposes in the spectrum is expected to provide an adequate estimation.}.
As we will argue below by inspecting the corresponding flux, the presence of Kolmogorov scaling implies an underlying cascade process in the droplet.

In contrast, for case II, we observe a $k^{-1}$ scaling among $k_L$ and $\xi^{-1}$. 
This is traced back to the formation of a random vortex distribution throughout the droplet background. 
It is also in line with previous  predictions in the repulsive interaction regime where it is known that this scaling behavior is generally associated with a uniform random vortex distribution in 2D repulsive gases~\cite{Estrada:PRA2022, Estrada:AVSQS2022} and with Vinen turbulence in 3D systems~\cite{Baggaley:PRB2012}. 
Turning to case III, we find that the droplet system  shows once again $k^{-5/3}$ Kolmogorov scaling which extends across a decade of length scales, i.e. from  $k_L$ to $\xi^{-1}$. 
The appearance of the Kolmogorov scaling is attributed to the underlying vortex turbulence as in case I. 
However, it is worth noting that the momentum range of this Kolmogorov scaling is much larger compared to case I which originates from the formation of vortex clusters (case III) instead of vortex dipoles (case I) preceding the turbulent stage. 

\begin{figure*}[!t]
\centering
\includegraphics[width=\linewidth]{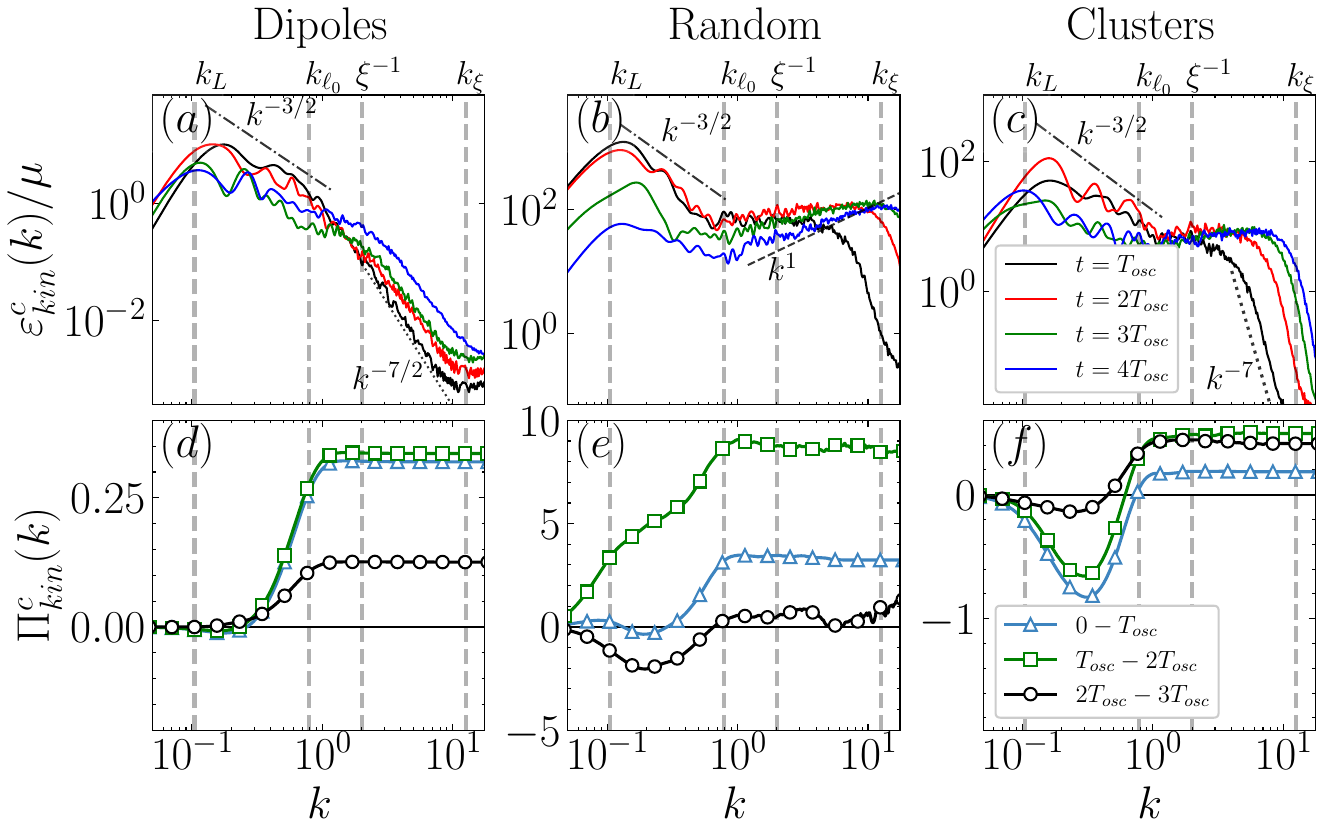}
\caption{(a)-(c) Compressible kinetic energy spectra and (d)-(f) associated fluxes of the driven 2D homogeneous droplet environment (with $N=10^4$ bosons) at the end of different oscillation periods, see legend. 
The stirrer has (a), (d) $(v,V_0) = (1.0,2.5)$ (case I), (b), (e) $(v, V_0) = (2.0,8.5)$ (case II) and (c), (f) $(v,V_0) =(1.0, 8.5)$ (case III). 
Momentum scales $k_L$, $k_{l_{0}}$ and $k_{\xi}$ depicted at the top of panels (a)-(c) are the same with the ones in Fig.~\ref{fig:EKI_spec_uniform}.   
All three cases exhibit $k^{-3/2}$ scaling in the infrared ($k<k_{l_0}$) range. 
This scaling signifies the presence of weak wave turbulence in the droplet type background. 
In the ultraviolet ($k>k_{l_0}$) range, case I shows a $k^{-7/2}$ scaling, whereas case II exhibits a tendency to approach $k$ scaling. The latter suggests a trend towards thermalization and its effective range in terms of length scales appears to expand over time. For case III a $k^{-7}$ decay is observed in the ultraviolet. The magnitude of the  corresponding flux, in general, increases from the first to the second oscillation period and substantially reduces in the third period, see also the black line signifying zero flux. }\label{fig:EKC_spec_uniform}
\end{figure*}

The underlying energy fluxes  [Eq.~(\ref{eq:flux_def})] of the incompressible kinetic energy for the cases I, II and III are illustrated in Fig.~\ref{fig:EKI_spec_uniform}(d)-(f). 
These were averaged in $\Delta t = T_{{\rm osc} }$ time-intervals so as to explicate the net energy transfer during each rotation cycle of the obstacle. 
In all three cases, it is observed that the first two cycles produce noticeable positive flux at all length scales followed by a significantly reduced  flux in the course of the third rotation taking even negative values at specific time-intervals. However, the magnitude and behavior of the fluxes differ for the distinct rotation cycles. 
This is traced back to both the characteristics of the driving and the nature of emergent defect configuration as we will explain in what follows.

Particularly, for case I where vortex dipoles emerge the maximum $\Pi^i_{{\rm kin} }(k)$ occurs for $k_L<k<k_{\ell_0}$ while $\Pi^i_{{\rm kin} }(k)$ is in general larger during the second rotation cycle and reduced during the third one. 
The fact that within the first two cycles $\Pi^i_{{\rm kin} }(k)>0$ suggests an accumulation of incompressible kinetic energy being more prominent at intermediate length scales where vortices form. 
Moreover, the amount of incompressible kinetic energy is directly linked to the amount of nucleated vortices~\cite{Bradley2012}. Hence, as the stirrer introduces more vortices into the system from the first to the second rotation cycle $\varepsilon^{i}_{{\rm kin} }(k)$ increases with time at all length scales and it is associated with a net positive flux. 
In contrast, $\Pi^i_{{\rm kin} }(k)$ is reduced during the third stirring cycle where the stirrer is slowly removed and vortex dipoles annihilate. 
Momentum regions with $\Pi^i_{{\rm kin} }(k)<0$ are likely to signify dominance of vortex-antivortex annihilation processes in the system. 
Finally, at large wavenumbers $k>\xi^{-1}$ the flux shows a constant behavior whose value depends on the considered time-interval. 
This is related to the observed $k^{-3}$ scaling of the incompressible kinetic energy [Fig.~\ref{fig:EKI_spec_uniform}(a)] for length scales smaller than the intervortex distance. 
Similar energy flux features are also evident for cases II and III. 
However, here the flux is almost two orders of magnitude larger than in case I. 
This is due to the formation of random vortex configurations and enhanced sound-waves in case II as well as vortex clusters in case III contrary to the vortex dipoles of case I. Another distinct feature of $\Pi^i_{{\rm kin} }(k)$ in cases II and III is that it appears to be nearly constant for $k_{\ell_0}$ as opposed to case I where it becomes constant only after $\xi^{-1}$. 
This hints towards the fact that  $\varepsilon^i_{{\rm kin} }$ shows power-law scaling for a wider range of length scales in the infrared region in cases II and III [Fig.~\ref{fig:EKI_spec_uniform}(b), (c)] and it is inspired by results on incompressible classical turbulence~\cite{Boffetta:ARFM2012}. However, one should notice that the current setup is more complicated since energy transfer might not be solely  restricted among different wavenumbers but also account for intercomponent energy contributions, compressible ones  as well as the LHY term.

\subsection{Compressible Spectra and Fluxes}\label{sec:fluxes_compressible}

Having analyzed the behavior of the incompressible kinetic energy spectra related to the occurrence of vortex turbulence we next discuss, for completeness, the respective compressible kinetic energy spectra, see also Eq.~(\ref{eq:spectra_def}). Figure~\ref{fig:EKC_spec_uniform}(a)-(c) presents the underlying compressible kinetic energy spectra at different time-intervals and for all distinct cases I, II and III. 
We observe that for all setups a clear $k^{-3/2}$ scaling arises at large length scales or equivalently in the momentum interval [$k_L$, $k_{l_0}$]. 
This scaling is associated with a weak wave cascade in the system~\cite{Nazarenko:PDNP2006}. It evinces the inevitable  presence of sound waves due to stirring and the emergent vortex annihilation processes irrespectively of the driving properties. 
It worths to be mentioned at this point that overall the magnitude of the compressible energy spectra [Fig.~\ref{fig:EKC_spec_uniform}(a), (c)] is significantly smaller for cases I and III from the incompressible ones [Fig.~\ref{fig:EKI_spec_uniform}(a), (c)] due to the dominance of vortex turbulence in the system. 
This is in contrast to case II where the compressible [Fig.~\ref{fig:EKC_spec_uniform}(b)] and incompressible [Fig.~\ref{fig:EKI_spec_uniform}(b)] kinetic energy spectra are of the same order of magnitude manifesting the competition between vortex and wave turbulence in this scenario. 
However, at large wavenumbers $k>\xi^{-1}$ (or small length scales) the compressible spectra exhibit distinct scaling which strongly depends on the driving characteristics. 

Specifically, within case I the spectra at large wavenumbers ($k>\xi^{-1}$) show a $k^{-7/2}$ scaling at early evolution times. 
While this scaling exponent at large wavenumbers decreases during the evolution it is still different from linear (i.e., $k^1$) which would correspond to a thermalized state~\cite{Shukla:NJP2013,Numasato:PRA2010}. The latter means, in this context, that the energy has spread among all available modes. 
Contrary to this, in case II, there is a rapid convergence towards thermalisation at large wavenumbers as the spectra feature a power law scaling of $\sim k$ at a wide range of length scales, namely $k>k_{l_0}$, especially for longer evolution times. 
We remark here that for weak wave turbulence a scaling $\sim k^{-7/2}$ is anticipated~\cite{Nazarenko:PDNP2006,Reeves:PRA2012} at large momenta instead of $k$. The latter is a consequence of the decaying turbulence observed in our setup since energy is injected into the system solely during the stirring process and not continuously. 
Finally, in case III we observe at large wavenumbers a power law scaling $k^{-7}$ which is modified as time evolves but again never becomes linear. 
This demonstrates that the system remains to be far-from-equilibrium at least for the evolution times that we have checked.

The energy flux emanating from the above-discussed compressible kinetic energies is provided in Fig.~\ref{fig:EKC_spec_uniform}(d)-(f), for the different cases I, II and III. 
Focusing on case I, we find a positive flux especially at large wavenumbers. 
The flux is nearly equal during the first two stirring periods but decreases by more than half in the course of the third period and becomes suppressed afterwards. 
Turning to case II, the flux appears again to be positive for the first two stirring periods having a larger magnitude in the second period and attaining a constant value for $k>k_{l_0}$. 
Note here that the flux of the compressible energy [Fig.~\ref{fig:EKC_spec_uniform}(b)] is larger in magnitude than the one of the incompressible kinetic energy [Fig.~\ref{fig:EKI_spec_uniform}(b)] demonstrating again the prevalence of weak wave turbulence.
During the third period it significantly diminishes becoming even negative for $k<k_{l_0}$. 
For later times it is nearly zero. 
Finally, for case III the flux is negative for $k<k_{l_0}$ and positive for $k>k_{l_0}$. 
The negative flux in the low wavenumber regime ($k\lesssim k_{l_0}$) suggests an inverse cascade of the compressible kinetic energy which may be attributed to the generation of directional sound wave propagation as a result of the turbulent wave mixing.

\begin{figure}[!t]  
\centering
\includegraphics[width=\linewidth]{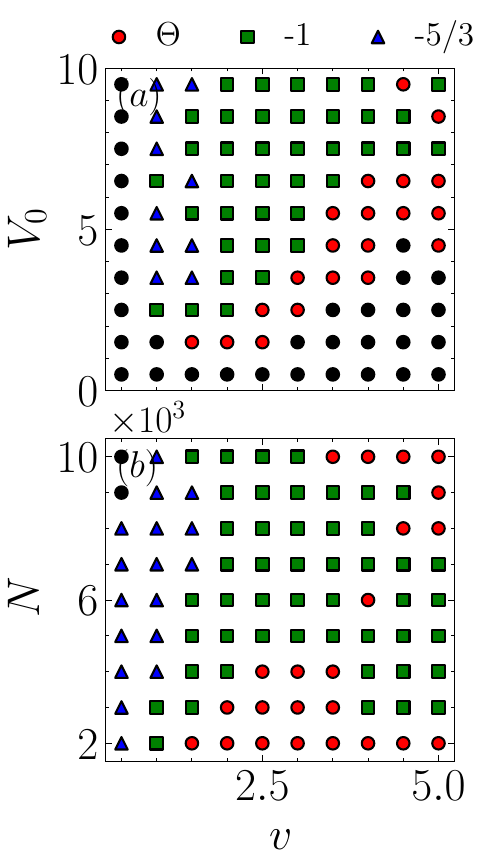}
\caption{Phase diagram demonstrating the scaling exponent (see legend) at small wavenumbers $k< \xi^{-1}$ (infrared) of the incompressible kinetic energy spectra $\varepsilon_{{\rm kin} }^{i}(k)$ at the end of the third stirring period of the homogeneous 2D  droplet environment. 
The resulting scaling exponent is presented as a function of (a) the potential height ($V_0$) and angular velocity $v$, as well as (b) the atom number $N$ and the velocity of the stirring potential. 
Green (blue) markers correspond to cases where a scaling $\sim k^{-1}$ ($\sim k^{-5/3}$) has been identified. 
Black dots represent parametric regions where vortex generation is absent at the end of the third stirring period, while red dots (indicated by the $\theta$ symbol in the legend) mark regimes at which it was not possible to assign a clean power law scaling in the infrared regime after the third period.  
The droplet atom number in panel (a) is $N=10^4$, while the barrier height in panel (b) is $V_0 = 5.5$.}\label{fig:phase_diagram_N_vs_Tosc}
\end{figure}

\subsection{Scaling dependence on the atom number and barrier characteristics} 

It becomes apparent from the above discussion that the interplay of the characteristics of the stirrer ($V_0$, $v$) and the atom number ($N$) are crucial for the different scaling behaviors of the driven droplet environment in the infrared regime.  
Hence, it is instructive to subsequently examine more carefully the respective scaling exponents for different parametric variations which will allow to identify the distinct types of emergent turbulent response. 
We remark that according to our simulations keeping fixed the barrier height, e.g. to $V_0=40$, and angular velocity ($v=\pi/4$), while considering different $N$ values we were able to identify the persistence of the above-discussed scalings of the energy spectra. Namely, i) $\varepsilon_{{\rm kin} }^{i}(k) \sim k^{-5/3}$ 
at $k<\xi^{-1}$ and as $\varepsilon_{{\rm kin} }^{i}(k) \sim k^{-3}$ for  $k>\xi^{-1}$, while ii) $\varepsilon_{{\rm kin} }^{c}(k) \sim k^{-3/2}$ at small wavenumbers\footnote{In fact, at large $N \geq 10^4$, the range of length scales showing $k^{-3/2}$ scaling decreases.}, i.e. $k>\xi^{-1}$ and $\varepsilon_{{\rm kin} }^{c}(k) \sim k^{-7/2}$ otherwise. 
However, since the stirrer facilitates the generation of vortices mainly captured via the incompressible kinetic energy, below we aim to focus on these spectra.

Figure~\ref{fig:phase_diagram_N_vs_Tosc} summarizes the identified scaling exponents of $\varepsilon_{{\rm kin} }^{i}(k)$ at intermediate length scales (infrared region, $k< \xi^{-1}$) with respect to the atom number, the height of the obstacle and its angular velocity. 
In all cases, the scaling is evaluated at the end of the third stirring period where defects (if any) have been fully formed. 
Focusing on the $V_0$-$v$ plane, we observe a somewhat complex dependence of the emergent scaling behavior on the barrier characteristics. 
Relatively small obstacle velocities, e.g. $v<1$, do not lead to vortex generation as indicated by the black boxes in Fig.~\ref{fig:phase_diagram_N_vs_Tosc}(a). This is attributed to the fact that the stirring velocity is lower than the critical one for the specific atom number to produce defects on top of the droplet background. 
On the other hand, an increasing angular velocity in the range $1 \leq  v<2$ leads to either a Kolmogorov scaling ($k^{-5/3}$ marked by the blue triangles) or random vortex distribution ($k^{-1}$ scaling indicated by the green squares) depending on the obstacle height, $V_0$. 
As expected, the occurrence of a random vortex distribution proliferates for larger velocities. 
Interestingly, there is a parametric region of increasing velocity ($v \geq 3$) and obstacle height ($V_0<3$) which do not lead to vortex generation after the first stirring period, see the black boxes circles in Fig.~\ref{fig:phase_diagram_N_vs_Tosc}(a). Additionally, there is a non-monotonous ($V_0-v$) regime at which a clear scaling behavior of $\varepsilon_{{\rm kin} }^{i}(k)$ is absent (red circles in Fig.~\ref{fig:phase_diagram_N_vs_Tosc}(a)).  
A similar but admittedly less complicated dependence of $\varepsilon_{{\rm kin} }^{i}(k)$ takes place as a function of the atom number and barrier height, see Fig.~\ref{fig:phase_diagram_N_vs_Tosc}(b). 
Here, smaller (larger) velocities in general facilitate
kolmogorov scaling (random vortex configurations) with the particle number variation seemingly playing a less major role.

\section{Vortex turbulence in a flat-top droplet}\label{turbulence_FT}

Next, we consider the turbulent response of a stirred droplet which possesses a flat-top shape. 
To achieve this in a controlled manner a weak 2D harmonic trap is assumed with angular frequency $\omega=0.003$, namely $V_{{\rm trap}} = \omega^2(x^2 + y^2)/2$. 
The system consists of $N=2 \times 10^4$ atoms in the  presence of an obstacle given by Eq.~(\ref{eq:Vobs}) with height $V_0=0.1$  residing at positions $x_0(0)=50$ and $y_0(0)=0$. 
These parameters ensure that the obstacle lies within the droplet background and in fact the ground state of the respective eGPE\footnote{Here, the flat-top ground state for $N=2 \times 10^4$ is realized with a box size of $320\times 320$.} 
supports a flat-top droplet solution featuring a density dip in the vicinity of the obstacle. 
Having obtained the droplet's ground-state, we next study its dynamical response via letting the gaussian obstacle of height $V_0=0.1$ and width $\sigma_0=3$ described by Eq.~(\ref{eq:Vobs}) to rotate. This triggers vortex formation and subsequently drives the system into the turbulent regime.  

\begin{figure*}[!t]
\centering
\includegraphics[width=\textwidth]{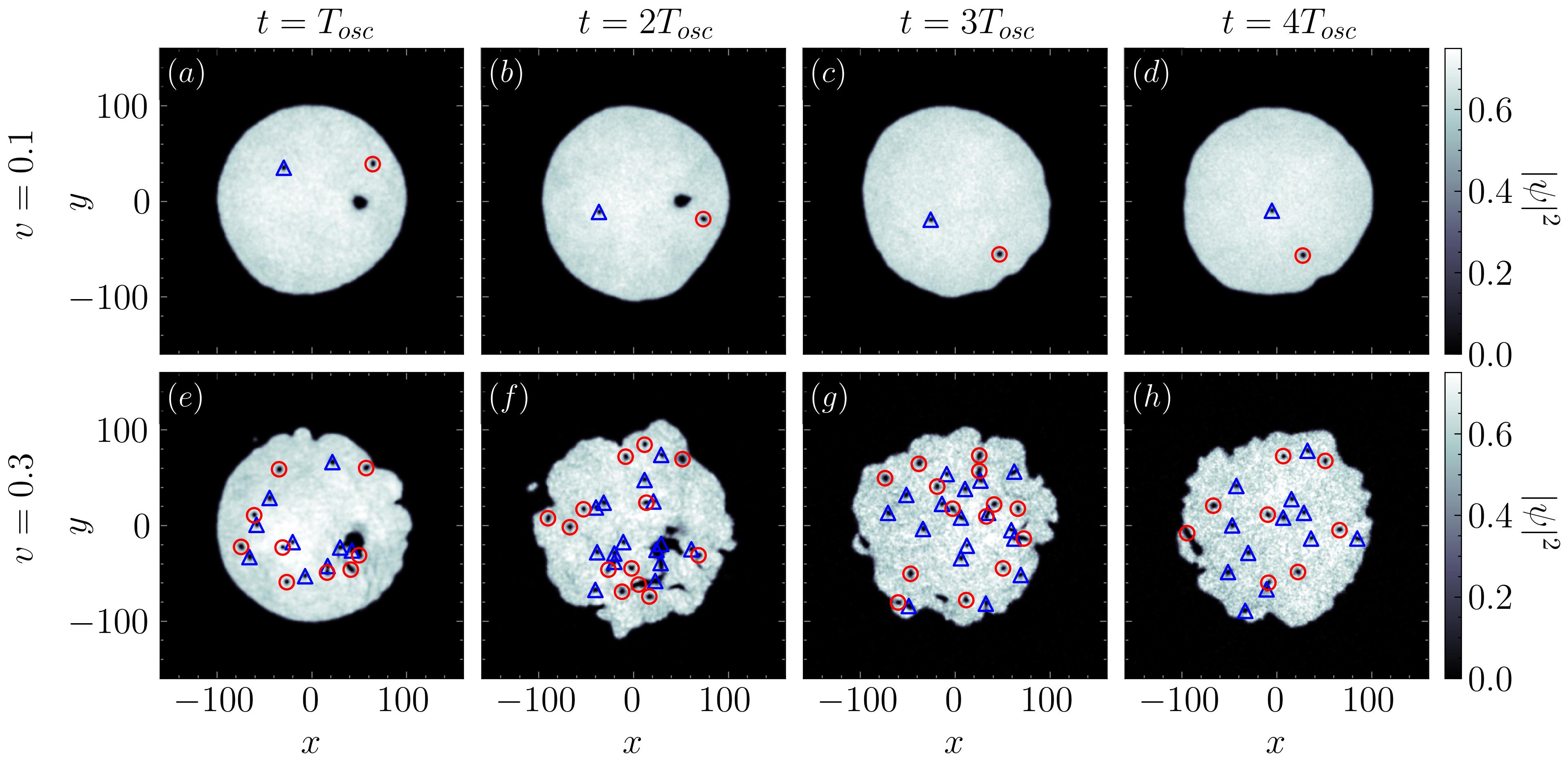}
\caption{Density profiles at different stirring periods (see legends) 
of a 2D flat-top quantum droplet. 
The rotating potential of width $\sigma_0=3$ is characterized by (a)-(d) $(V_0,v) = (0.1,0.1)$, and (e)-(h) $(V_0, v) = (0.1, 0.3)$. 
The potential is removed after the third oscillation period (panels (c), (g)). 
It is evident that for lower stirring velocities [panels (a)-(d)] a vortex dipole is created, while for increasing velocities a multitude of vortex-antivortex pairs [panels (e)-(h)] occurs on top of the droplet background featuring annihilation events during the evolution. 
Vortices (antivortices) are designated by the red circles (blue triangles). 
In all cases, the droplet consists of  $N=2 \times 10^4$ bosons and it is under the influence of a 2D harmonic trap of radial frequency $\omega=0.003$. }\label{fig:density_flattop}
\end{figure*}

Density profiles of the perturbed flat-top droplet at different evolution times are depicted in  Fig.~\ref{fig:density_flattop} for stirring velocities $v=0.1$ [Fig.~\ref{fig:density_flattop}(a)-(d)] and $v=0.3$ [Fig.~\ref{fig:density_flattop}(e)-(h)]. 
In the first case, with relatively smaller stirring velocity, the gaussian potential seeds the creation of a vortex-antivortex pair in the flat-top background, see Fig.~\ref{fig:density_flattop}(a)-(b). 
The existence of the vortex and the anti-vortex have been confirmed by inspecting the respective phase of the time-dependent droplet wave function exhibiting a clockwise (counter-clockwise) $2\pi$ phase-jump across the vortex (antivortex) core, not shown for brevity. 
This vortex dipole rotates following the movement of the stirring potential, with the vortex and antivortex precessing around the droplet core in opposite directions. Moreover, their distance changes in the course of the evolution since they appear to affect its others motion by means that their  precession rate slows down when they come closer. 
Simultaneously, the periphery of the droplet becomes slightly distorted which is attributed to the emission of  stirring-induced density disturbances (involving sound-waves) towards the droplet edges (hardly visible in the densities). 
Finally, the vortex dipole remains trapped within the droplet even after the removal of the barrier at the end of the third oscillation period and in fact persists (i.e. does not escape) for long evolution times that we have checked, i.e. $t \sim 6 \times 10^4$ in dimensionless units.  

Turning to larger stirring velocities, the droplet response appears to be drastically altered, see Fig.~\ref{fig:density_flattop}(e)-(h). 
Namely, the production of a large number of vortex-antivortex pairs takes place into the droplet background and hence vortex clustering is observed. 
This process is accompanied by an appreciable amount of emitted density distortions in the form of sound-waves due to the rotation of the gaussian potential but also vortices which travel towards the droplet boundary and escape from it. 
These sound-waves and boundary vortices impact  the integrity of the droplet to a non-negligible degree, unlike the case with $v=0.1$ presented in Fig.~\ref{fig:density_flattop}(a)-(d), by means of substantially deforming the droplet boundary which becomes quite irregular. 
Additionally, some of the vortex and antivortex entities come closer to each other and get annihilated as time-evolves. 
In this way, the number of vortex-antivortex pairs reduces during the dynamics and for instance five vortices and one antivortex remain for long evolution times, e.g. $t=6 \times 10^4$.

We remark that upon considering larger barrier heights such as twice the previous one (i.e. $V_0=0.2$), while maintaining similar velocities, e.g. $v=0.2$ or 0.3 the rotating obstacle sheds an irregular vortex distribution in its wake but not vortex pairs as in the $(V_0,v)=(0.1, 0.3)$ case. 
Here, we find that the integrity of the droplet is even more  substantially affected by the stirring process. 
Finally, utilizing even larger values of $V_0$ and $v$ leads to breaking of the droplet. 
A process that is reminiscent of the granulation phenomenon  which has been observed both experimentally~\cite{Seman:LPL2011,Nguyen_granulation} and numerically~\cite{Yukalov:JLTP2015} in scalar Bose gases when large amplitude excitations are at play. 
This mechanism for quantum droplets remains elusive and constitutes an intriguing prospect for future studies.  
It should be noted, however, that larger velocities translate to pumping a significant amount of energy into the droplet. 
When this injected energy becomes comparable to the droplet's binding energy, the droplet will break apart, which is the self-evaporation phenomenon~\cite{petrov2016ultradilute}.

\begin{figure}[!t]
\centering
\includegraphics[width=1.0\linewidth]{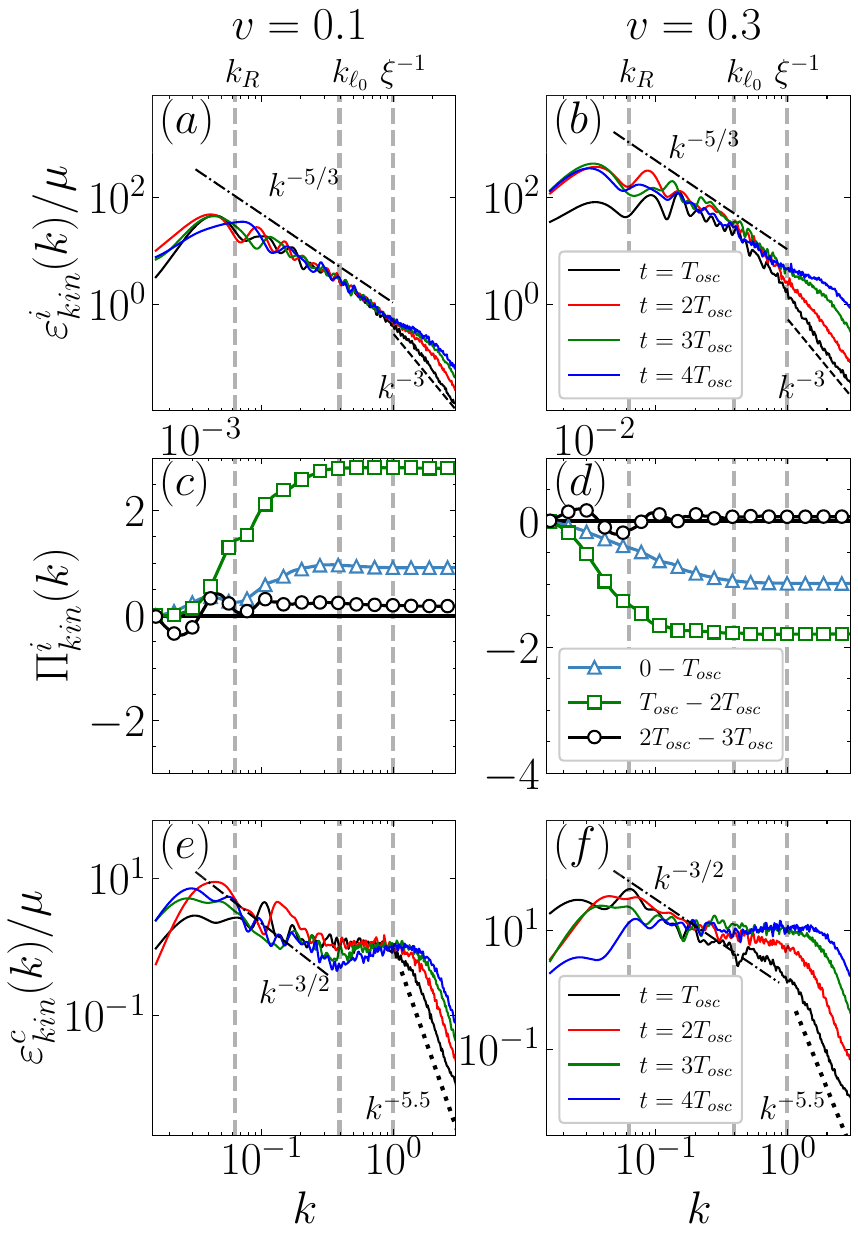}
\caption{(a), (b) Incompressible kinetic energy spectra, (c), (d) energy flux and (e), (f) compressible kinetic energy spectra of the 2D harmonically trapped droplet with $N=2 \times 10^4$ subjected to a stirring potential. 
The latter has (a), (c), (e) $(V_0,v) = (0.1,0.1)$ and (b), (d), (f) $(V_0, v) = (0.1, 0.3)$, see also Fig.~\ref{fig:density_flattop}. 
$k_R$ and $k_{l_{0}}$ refer to the momentum scales of the droplet and mean intervortex distance, with $\mu$ being the chemical potential of the 2D droplet.  
The black solid lines in panels (d)-(f) mark zero flux. 
For $\varepsilon_{{\rm kin} }^{i}(k)$, we observe Kolmogorov scaling, $k^{-5/3}$, in the infrared regime ($k<\xi^{-1}$), while $k^{-3}$ scaling occurs in the ultraviolet ($k>\xi^{-1}$) regime independently of the driving characteristics.    
Similarly, $k^{-3/2}$ scaling takes place in the infrared for $\varepsilon_{{\rm kin} }^{c}(k)$, and $k^{-5.5}$ in the ultraviolet. 
The respective incompressible energy flux is positive (negative) for smaller (larger) velocities of the obstacle.}
\label{fig:eki_spectra_flattop}
\end{figure}

To further understand the aforementioned nonequilibrium behavior of the driven flat-top droplet we resort to the corresponding kinetic energy spectra. 
In particular, the incompressible kinetic energy spectra for $V_0=0.1$ with either $v=0.1$ or $v=0.3$ are provided in Fig.~\ref{fig:eki_spectra_flattop}(a), (b) respectively. 
It can be readily seen that in both cases $\varepsilon_{{\rm kin} }^{i}(k)$ exhibit a Kolmogorov  scaling in the infrared regime spanning a decade of wavenumbers from $k_R$ to $\xi^{-1}$. Here, $k_R$  represents the momentum scale associated with the droplet radius being for our setup $R\approx 100$ in the adopted dimensionaless units. 
On the other hand, in the ultraviolet regime i.e., for $k \gtrsim \xi^{-1}$ we find that $\varepsilon_{{\rm kin} }^{i}(k) \sim k^{-3}$. 
This scaling is characteristic of the vortex core structure as discussed in Sec.~\ref{sec:fluxes_homogeneous}. 
Note here the similar scaling of $\varepsilon_{{\rm kin} }^{i}(k)$ in both the infrared and ultraviolet regimes with  the homogeneous driven droplet bearing environment featuring vortex turbulence, see also  Fig.~\ref{fig:EKI_spec_uniform}(a), (b). 
Moreover, it is interesting to remark that $\varepsilon_{{\rm kin} }^{i}(k)$ does not show a tendency towards a self-similar behavior in the ultraviolet regime as observed for the homogeneous setup [Fig.~\ref{fig:EKI_spec_uniform}(a)-(c)] for long evolution times.

For completeness, we also present in both cases the associated energy flux of the incompressible kinetic energy in Fig.~\ref{fig:eki_spectra_flattop}(c), (d). 
These are averaged within consecutive oscillation periods in order to capture the respective energy transfer. 
In both cases, we can deduce that the magnitude of the flux increases from the first to the second driving period and it is reduced in the third rotation cycle. 
The enhancement from the first to the second period is traced back to the accompanied amplification of sound-waves for $v=0.1$ and accumulation of vortices for $v=0.3$.  
Additionally, the flux decreases in the third stirring period because the potential well is removed and vortex dipoles annihilate. 
However, the sign of the fluxes are reversed for increasing angular velocity of the obstacle. 
This reversion is attributed to the pronounced vortex annihilation processes in the case of larger angular  velocity. 
Finally, for length scales larger than the intervortex distance, i.e. $k>k_{l_0}$, the flux attains a constant value. 

Since the stirring process produces a non-negligible amount of sound-waves especially for larger rotation velocities it is natural to also examine the compressible kinetic energy spectra, $\varepsilon_{{\rm kin} }^{c}(k)$. Figures~\ref{fig:eki_spectra_flattop}(e), (f) illustrate $\varepsilon_{{\rm kin} }^{c}(k)$ for both driving scenaria discussed above. 
Irrespectively of the driving characteristics the spectrum  shows a $k^{-3/2}$ scaling at small wavenumbers ($k<k_{l_0}$) which evidences the existence of a weak wave turbulent cascade in the system. 
Additionally, we find that  $\varepsilon_{{\rm kin} }^{c}(k) \sim k^{-5.5}$ in the ultraviolet regime, which does not allude to a known scaling and hence deserves further investigation in future studies.
It is also worth noting that the magnitude of the  compressible kinetic energy is larger for increasing angular velocity, compare panels (e) and (f) in  Fig.~\ref{fig:eki_spectra_flattop}.  
This is because a faster obstacle induces a larger amount of sound waves and more prominent vortex-antivortex annihilations throughout the evolution.

\section{Conclusions and perspectives}\label{sec:con}

We have investigated the emergent turbulent response of 2D quantum droplet environments utilizing a stirring optical potential which facilitates the formation of vortices and sound-waves. 
To treat the ensuing nonequilibrium quantum dynamics of droplets we employed the appropriate 2D eGPE characterized by a logarithmic nonlinear term which encompasses mean-field interactions and the first-order LHY quantum correction. 
While our main focus is placed on a 2D homogeneous box trapped droplet environment, we generalize our results to a harmonically trapped flat-top droplet in the presence of a static potential barrier (stirrer). 
It is shown that in the box potential, an increasing atom number results in the deformation from a gaussian type to a flat-top droplet and eventually leads to a homogeneous background as a consequence of the finite box size. 
The optical barrier, when placed within the droplet, appears as a density notch.   

The dynamics is triggered upon stirring the optical barrier inside the droplet environment for three full oscillation periods and afterwards removing it, while letting the system to further evolve. 
The resultant dynamical response is classified based on the nature of the density defects, the second-order sign correlation function and importantly on the spectra of the underlying incompressible and compressible kinetic energies as well as their associate fluxes. 
It is exemplified that the turbulent behavior depends crucially on the characteristics (height and velocity) of the stirring potential. 

Specifically, for relatively small velocities and height of the barrier vortex dipoles (case I) occur, while an increasing height leads to the formation of randomly distributed vortex clusters accompanied by  an enhanced amount of sound-waves (case II). 
A further increase of the stirring velocity at relatively large heights results in the generation of vortex-antivortex clustering (case III).  
In all of these settings, the incompressible kinetic energy spectra exhibit a $k^{-3}$ scaling in the ultraviolet regime emanating from the presence of vortices, while showing a self-similar behavior at long evolution times. 
On the other hand, within the infrared range cases I and III feature Kolmogorov $k^{-5/3}$ scaling due to vortex turbulence, whilst case II presents a $k^{-1}$ scaling traced back to the formation of the random vortex distribution. The compressible spectra, in all three cases, scale as $k^{-3/2}$ in the infrared regime suggesting the presence of weak wave turbulence.  
In the ultraviolet, case I presents a $k^{-7/2}$ scaling, and cases II and III have a trend towards a $k$ scaling signaling thermalization. 
Overall, the respective energy fluxes increase from the first to second oscillation period and subsequently decrease in the third one since stirring terminates. 

Turning to a flat-top droplet background we showcase that for an increasing stirring velocity the dynamical response transits from a vortex-dipole dominated regime to one where an appreciable amount of vortex-antivortex pairs occurs. 
Here, the droplet periphery suffers significant distortions especially when more vortex-antivortex pairs are present due to enhanced density disturbances including sound waves and vortices traveling to the droplet edges. 
Similarly to the homogeneous droplet environment, the incompressible kinetic energy spectra (in all studied cases) exhibit a Kolmogorov $k^{-5/3}$ scaling in the infrared regime, and a $k^{-3}$ scaling characteristic of the vortex core in the ultraviolet. 
Furthermore, the compressible kinetic energy spectra, feature irrespectively of the driving characteristics a power law  $k^{-3/2}$ in the infrared evincing the involvement of a weak wave turbulent cascade. 

Our results provide the starting point for a multitude of intriguing future research  directions based on turbulent response. 
An interesting extension is to deploy other driving protocols in order to trigger turbulence in relevant three-dimensional droplet settings, e.g. by crossing the droplet-to-gas transition~\cite{Semeghini2018}, or examining the emergent cascades and related scalings of energy spectra in the crossover from three- to two-dimensions. 
Furthermore, studying the transition from turbulence to granulation using, for instance, larger amplitude perturbations and/or velocities of the stirrer is another interesting extension of our findings. 
Also, the design of appropriate protocols to trigger inverse energy cascades in droplets, e.g. via periodic external potentials~\cite{Reeves2013,karailiev2024observation} is of immense interest. Moreover, proceeding a step forward in order to characterize strongly interacting turbulence as was recently done in Ref.~\cite{zhu2024transition} in the context of Bose gases or by resorting to beyond eGPE numerical techniques~\cite{mistakidis2023few,cao2017unified} allowing the exploration of the ensuing  correlation patterns would be worth pursuing.

\section*{Acknowledgements} 
S.K.J gratefully acknowledges financial support from the University Grants Commission - Council of Scientific and Industrial Research (UGC-CSIR), India.~S.I.M acknowledges support from the Missouri Science and Technology, Department of Physics,
Startup fund.~S.I.M thanks G.~Bougas and  K. Mukherjee for
fruitful discussions on the topic of turbulence. M.K.V. gratefully
acknowledges the support from Science and Engineering Research Board, India, for the J. C. Bose Fellowship (Grant No. SERB/PHY/2023488) and for Grants No.
SERB/PHY/2021522 and No. SERB/PHY/2021473. We acknowledge the fruitful discussions with Sachin S. Rawat at initial stage of the work.

\appendix

\section{Scaling at large momenta of a vortex in a scalar BEC and a droplet}\label{scaling_vortex}

\begin{figure}[!t]
\centering
\includegraphics[width=1.0\linewidth]{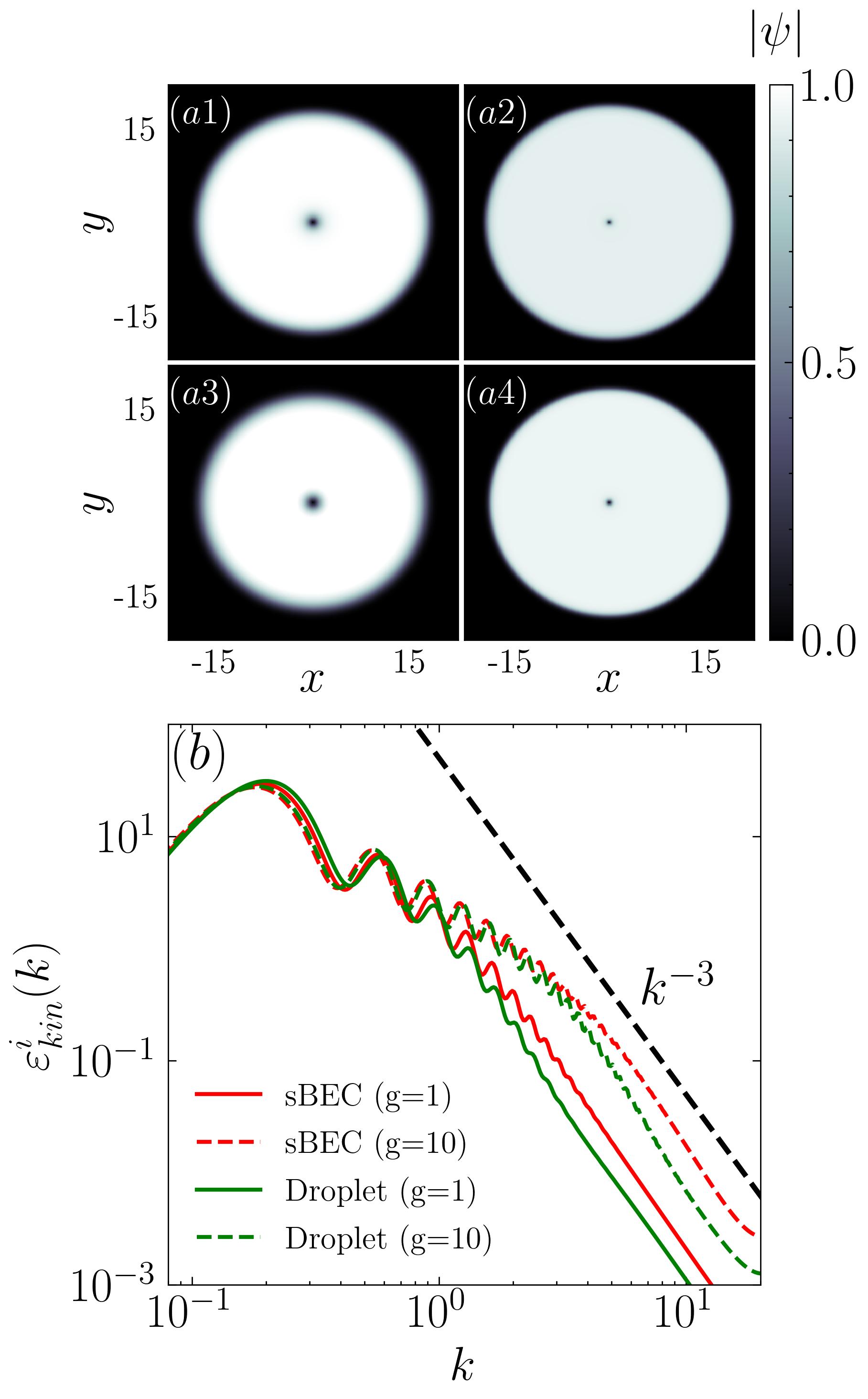}
\caption{Amplitude of the 2D wave function for the entire system of (a1), (a2) a scalar BEC indicated by sBEC and (a3), (a4) a 2D droplet  with (a1), (a3) $g=1$, and (a2), (a4) $g=10$. 
The vortex is generated in the bulk through phase imprinting in the absence of any obstacle. 
(b) The corresponding incompressible kinetic energy spectra in all different cases (see legend) feature a $k^{-3}$ scaling in the ultraviolet regime. The considered atom number is $N=900$.}
\label{fig:SVS}
\end{figure}

In the main text, we argued that the incompressible kinetic energy spectra of a driven homogeneous droplet environment in the ultraviolet regime exhibit a $k^{-3}$ scaling behavior which is attributed to the presence of vortices. Here, we aim to demonstrate that this scaling indeed appears when a vortex is embedded either in a 2D scalar BEC or a quantum droplet. Moreover, this study will exemplify differences between the incompressible spectra of vortices in a scalar BEC and a  droplet environment. 
Both BEC and droplet backgrounds contain $N=900$ bosons being trapped in a 2D circular  box potential of size $L \times L= 80 \time 80 $. 
The scalar BEC is treated within the well-known 2D GPE~\cite{pitaevskii2016bose}, while the droplet environment with the 2D eGPE~(\ref{eq:2Degpe}). 
We imprint a single vortex of unit charge at the center of the appropriate (scalar BEC or droplet) background confined by the following circular box trap  
\begin{equation}
V_{{\rm trap}} = \frac{U_0}{2}\left[\tanh(r-R_0)+1\right].
\end{equation} 
Here, $U_0 = 100$ refers to the height of the potential and $R_0=20$ is its effective radius. 
Moreover, the used ansatz to obtain the single vortex radial wave function in both settings is given by 
\begin{equation}
\phi_0(r) = C\,r\,\exp(-\alpha r^2 + i\theta), 
\end{equation} 
where $r=\sqrt{x^2+y^2}$ denotes the radial distance, $\theta=2 \pi$ is the azimuthal angle describing the vortex phase circulation, $\alpha=10^{-2}$ and $C$ is the normalization constant. 
With this initial condition, we perform  imaginary-time propagation of the underlying GPE or eGPE while keeping the phase fixed to its  initial value. 
This process leads to a unit charge vortex solution located at the center of the corresponding background. 
The state is assumed to be converged when the system's energy difference between two consecutive time steps in the imaginary time-propagation is below $10^{-8}$. 

Figure~\ref{fig:SVS} presents paradigmatic vortex densities in the cases of a scalar BEC [Fig.~\ref{fig:SVS}(a1), (a2)] and a quantum droplet environment [Fig.~\ref{fig:SVS}(a3), (a4)] for two different values of the involved nonlinear parameter, $g$. 
Recall that the coupling strength, $g$, in the GPE models repulsive mean-field interactions, while in the eGPE contains the effect of both the mean-field interactions and the LHY quantum fluctuations~\cite{petrov2016ultradilute}. 
In all cases, the density depletion at $(x,y)=(0,0)$ represents the vortex which is further characterized by a $2 \pi$ phase jump. 
Furthermore, independently of the model an increasing interaction value leads to a shrinked vortex core and a wider background, see for instance Fig.~\ref{fig:SVS}(a1), (a2). 
Also, the vortex core for fixed interactions is somewhat larger in the case of a droplet environment, e.g. compare Fig.~\ref{fig:SVS}(a1), (a3). 
On the other hand, the corresponding incompressible kinetic energy spectra of these vortex states are presented in Fig.~\ref{fig:SVS}(b). A clear $k^{-3}$ scaling is observed in the ultraviolet range ($k>\xi^{-1}$) for both systems irrespective of the strength of the nonlinearity and despite the difference in the vortex core size between them. 
For stronger interactions slight distortions of the spectra occur in the ultraviolet regime which is attributed to the fact that  the GPE models are valid in the weak interaction regime. 

\bibliography{references.bib}
\end{document}